\documentclass[%
 reprint,
 amsmath,amssymb,
 aps,
]{revtex4-2}

\RequirePackage{graphicx}
\usepackage{amsmath}
\usepackage{amsfonts}
\usepackage{amsthm}
\usepackage{mathrsfs}
\usepackage{amssymb}
\usepackage{epsfig}
\usepackage{hyperref}
\usepackage{color}
\usepackage{amscd}
\usepackage{xcolor}

\def\prd{Phys. Rev. D }

\def\mnras{Monthly Notices of the Royal Astronomical Society }
\def\apj{The Astrophysical Journal }
\def\aap{Astronomy and Astrophysics }

\def\pasj{Publications of the Astronomical Society of Japan }

\def\nat{Nature }

\newcommand{\beq}{\begin{equation}}
\newcommand{\eeq}{\end{equation}}
\newcommand{\bea}{\begin{eqnarray}}
\newcommand{\eea}{\end{eqnarray}}

\providecommand{\dif}{\mathrm{d}} \def\d{\dif}

\def\EE{{\cal E}}
\def\LL{{\cal L}}
\def\BB{{\cal B}}

\def\mJ{I} \def\mD{II}

\newcommand{\ce}{{\cal{E}}} 
\newcommand{\cl}{{\cal{L}}} 
\newcommand{\cb}{{\cal{B}}}

\graphicspath{{pic/}}

\def\Opava{Research Centre for Theoretical Physics and Astrophysics, Institute of Physics, Silesian University in Opava, CZ-74601 Opava, Czech Republic}
\def\Almaty{Al-Farabi Kazakh National University, Almaty 050040, Kazakhstan}
\def\Kazakh{Department of Physics, Kazakh National Women’s Teacher Training University, 050000 Almaty, Kazakhstan}
\def\Bonn{Max Planck Institute for Radio Astronomy, Auf dem H{\"u}gel 69, Bonn D-53121, Germany}

\begin{document}

\title{ Black hole in a combined magnetic field: ionized accretion disks in the jetlike and looplike configurations
}

\author{Saltanat Kenzhebayeva}\email{s.s.kenzhebayeva@gmail.com}\affiliation{\Almaty}
\author{Saken Toktarbay}\email{toktarbay.saken@kaznu.kz}\affiliation{\Almaty}\affiliation{\Kazakh}
\author{Arman Tursunov}\email{arman.tursunov@physics.slu.cz}\affiliation{\Bonn}\affiliation{\Opava}
\author{Martin Kolo\v{s}}\email{martin.kolos@physics.slu.cz}\affiliation{\Opava}

\begin{abstract} 
 
Magnetic fields surrounding black holes are responsible for various astrophysical phenomena related to accretion processes and relativistic jets. 
Depending on the source, the configuration of the field lines may differ significantly, affecting the trajectories of charged particles and the corresponding observables. Usually, the magnetic fields around black holes are modeled within a single source or current generating the field.   
However, magnetic fields can have more than a single origin, being a combination of different fields, such as, e.g., that of an accretion disk and external large-scale or Galactic ones. 
In this paper, we propose a combined magnetic field solution given by the superposition of the uniform and Blandford-Znajek split-monopole magnetic fields in a strong gravity regime of the Schwarzschild black hole. 
We show that when the combined magnetic field components are aligned, the resulting field is of a paraboloidal jetlike shape. Such a configuration is supported by relativistic jet observations and is often utilized in general relativistic magnetohydrodynamical simulations. 
In the opposite orientation of the two field components, we observe looplike field structures magnetically connecting the black hole with an accretion disk and the magnetic null points, which can be related to the regions of magnetic reconnection.
In the combined magnetic field configurations, we analyze the dynamics of charged particles, study their stability conditions, and find the locations of stable off-equatorial structures close to the symmetry axis. Finally, we consider an ionization of Keplerian accretion disk as a particular scenario of particle scattering. From the numerical experiments, we conclude that charged particles in the jetlike combination show a strong tendency to escape from the black hole, which is not observed in the case of individual fields. In contrast, the looplike combination supports accretion of charged particles into the black hole. 

\end{abstract}

\maketitle

\section{Introduction}

Magnetic fields in regions of strong gravity  play a crucial role in the explanation of various high-energy phenomena observed in black hole systems of different mass scales. Among these phenomena are e.g. the formation and collimation of relativistic jets \cite{Bla-Zna:1977:MNRAS:} observed in active galactic nuclei (AGNs) and other black hole systems, the generation of high-frequency quasiperiodic oscillations (QPOs) from microquasars and AGNs \cite{2022PASJ...74.1220S}, acceleration of ultrahigh-energy cosmic rays \cite{Tur-etal:2019:ApJ:}, among others. 

The strength of the magnetic field around compact objects can be estimated using different methods depending on the type of object and properties of the surrounding plasma. A typical magnitude of the magnetic field around supermassive black holes at the centers of many galaxies is estimated to be around $B \sim 10^4$G \cite{Daly:APJ:2019:}. This value refers mainly to AGNs with visible relativistic jets. In contrast to that, the Galactic center supermassive black hole Sgr~A* in a quiescent state is likely surrounded by an ordered magnetic field, which is significantly weaker than in AGNs, being of the order of $\sim 10 - 100$G  \cite{Eatough-etal:2013:Natur:,Eckart-etal:2017:FOP:}.  For stellar mass black holes, the strengths of the magnetic field can vary all the way from mG for isolated black holes inside the Galaxy up to the values of $10^8$G, typical for the non-transient binary systems such as microquasars \cite{Pio-etal:2020:arXiv:}. All these and other estimates indicate that the magnetic field around black holes of various masses in the astrophysical context is a test field, meaning that it cannot alter the spacetime geometry. This condition reads $B \ll B_{\rm G} = 10^{19} {\rm G} \times M_{\odot} / M$ \cite{Tur-Stu-Kol:2016:PHYSR4:}. 
A field that satisfies the test field condition, $B \ll B_G$, has a negligible effect on neutral particles. However, for charged particles, the corresponding Lorentz force influence can be considerably large or dominant, as in the case of elementary particles. For a relativistic particle of charge $q$ and mass $m$, the ratio of the Lorenz force to the gravitational force close to the black hole can be represented by the following dimensionless parameter 
\beq
{\cal B} =\frac{|q| G M B}{2 m c^4}.
\label{BB-param}\eeq
This quantity can be quite large, even for weak magnetic fields due to the large value of the specific charge $q/m$ \cite{2010PhRvD..82h4034F,Kol-Stu-Tur:2015:CLAQG:}. Therefore, in many cases, the influence of the magnetic field on the motion of charged particles cannot be neglected \cite{Kol-Sha-Tur:2023:EPJC:,Stu-etal:2020:Universe:}.

Many observed black hole candidates are surrounded by accretion disks of conducting plasma. Depending on their densities, shapes, dynamics, and stages of accretion, the magnetic fields generated by the plasma can have different configurations and strengths. Additionally, the magnetic field around a black hole can also be of an external origin, e.g. magnetic field of the Galaxy, or of a companion neutron star as in the case of a binary system. Hence, the final shape of the magnetic field can be somewhat more complicated, being a combination of more than one field.

For the simplest approximation to the problem of black hole magnetosphere, one can use the vacuum solutions of the Maxwell's equations in the curved background as a starting point. Since the Maxwell's equations are linear, one can combine two or more different solutions and create a new one. The first exact solution of Maxwell's equations in a curved background of a rotating black hole was obtained by Wald \cite{Wald:1974:PHYSR4:}. It represents a solution of the vacuum Maxwell equations for an asymptotically uniform magnetic field, in which a Kerr black hole is immersed. The simple and elegant form of the Wald solution allows one to study various effects of electromagnetic fields on a charged matter in the strong gravity regime analytically, although ignoring collective plasma effects. Therefore, although the Wald solution can serve as a good approximation to describe external magnetic fields around isolated black holes, it is not very suitable for the description of the black hole magnetospheres generated by plasma accretion disks. 
Shortly thereafter, Blandford and Znajek (BZ) in their seminal work \cite{Bla-Zna:1977:MNRAS:}, investigated the plasma-filled black hole magnetosphere described by force-free electrodynamics and formulated the electromagnetic energy extraction mechanism from rotating black holes. In this mechanism, the magnetosphere around a slowly rotating Kerr black hole is considered to be a split-monopole. In this configuration, the magnetic field lines in the southern and northern hemispheres go in opposing directions, which is where the difference between the monopole and split-monopole solutions lies. Therefore, the BZ split-monopole solution does not require the presence of magnetic monopole charges and does not violate the no-hair theorem. Similar magnetic field configurations around the black hole have been explored within relativistic magnetohydrodynamics (GRMHD) \cite{Tch-Nar-McK:2010:APJ:,Kol-Jan:2020:RAG:,Jan-Jam:2022:AAP:} or relativistic particle-in-cell (GRPIC) \cite{Par-Phi-Cer:2019:PRL:,Cri-etal:2021:AAP:} numerical simulations.

Despite this, the exact shape of the magnetosphere around astrophysical black holes has not yet been properly resolved. In realistic cases, one can expect that the final form of the magnetosphere will result in a combination of internal and external solutions. While the internal field is strongly connected to the accretion processes, the external field is independent of plasma properties in the black hole vicinity, which can be, e.g., of a Galactic magnetic field, or of a companion neutron star. Due to the linearity of Maxwell's equations of electrodynamics, the combined field solution can be described as a superposition of the two individual field potentials, so the resulting solution satisfies the Maxwell equations. 

In this paper, we study the effects of a combined magnetic field given by a superposition of uniform and BZ split-monopole solutions. As an exact solution of the Maxwell's equations in the curved Schwarzschild and Kerr black hole spacetimes, the combined magnetic field has clear advantages over the previously considered heuristic parabolic black hole magnetosphere model \cite{Nak-etal:2018:APJ:,Kol-Sha-Tur:2023:EPJC:}. In particular, we focus on the influence of resulting field configurations on the dynamics of charged matter. 
Close to the black hole, where the effects of accretion processes are strong, we assume that the magnetic field is dominated by the BZ split-monopole solution \cite{Bla-Zna:1977:MNRAS:}. At larger distances from the black hole, we describe the external field using the homogeneous Wald solution \cite{Wald:1974:PHYSR4:}. At large distances from the source of an external magnetic field, in a finite region of space, the field lines can be approximated to be nearly uniform.

Assuming an axial symmetry of the combined magnetic field, we show that depending on the relative orientation of the components, the resulting field can be of a paraboloidal "jetlike" or of closed "looplike" shape. The former case is well supported by numerical simulations and observations of relativistic black hole jets. Observational studies of the collimation of relativistic jets in M87 give evidence of the parabolic shape of the jet closer to the formation zone and transiting to the conical shape further away from the black hole \cite{Nak-etal:2018:APJ:}. 
On the other hand, in the looplike configuration, the magnetic field lines provide a direct magnetic connection between a black hole and an accretion disk \cite{Cha-Abo:2015:MNRAS:}, which can restrict particle escape and support the accretion. In the looplike configuration, one can observe the existence of magnetic null-points, which are essential for magnetic reconnection. 

We organize the paper as follows. In Section \ref{sec:1} we introduce the combined
magnetic field configuration in a spherically symmetric Schwarzschild spacetime and discuss the properties of the new solutions. We show that the two solutions can be combined in two qualitatively different ways, which reminds either jetlike or looplike structures. In Section \ref{sec:2}, we study the dynamics of charged particles, analyze the effective potential, introduce equatorial and off-equatorial orbits, and find the position of the innermost stable orbit. In Section \ref{sec:Disk} we consider the ionization of a Keplerian accretion disk slightly tilted with respect to the magnetic field lines. We show that the jetlike configuration supports the escape of charged particles, while the looplike configuration supports particle accretion into the black hole. Finally, we give our concluding remarks and discuss the prospects of combined field studies.  

\begin{figure*}
\centering
\includegraphics[width=0.75\textwidth]{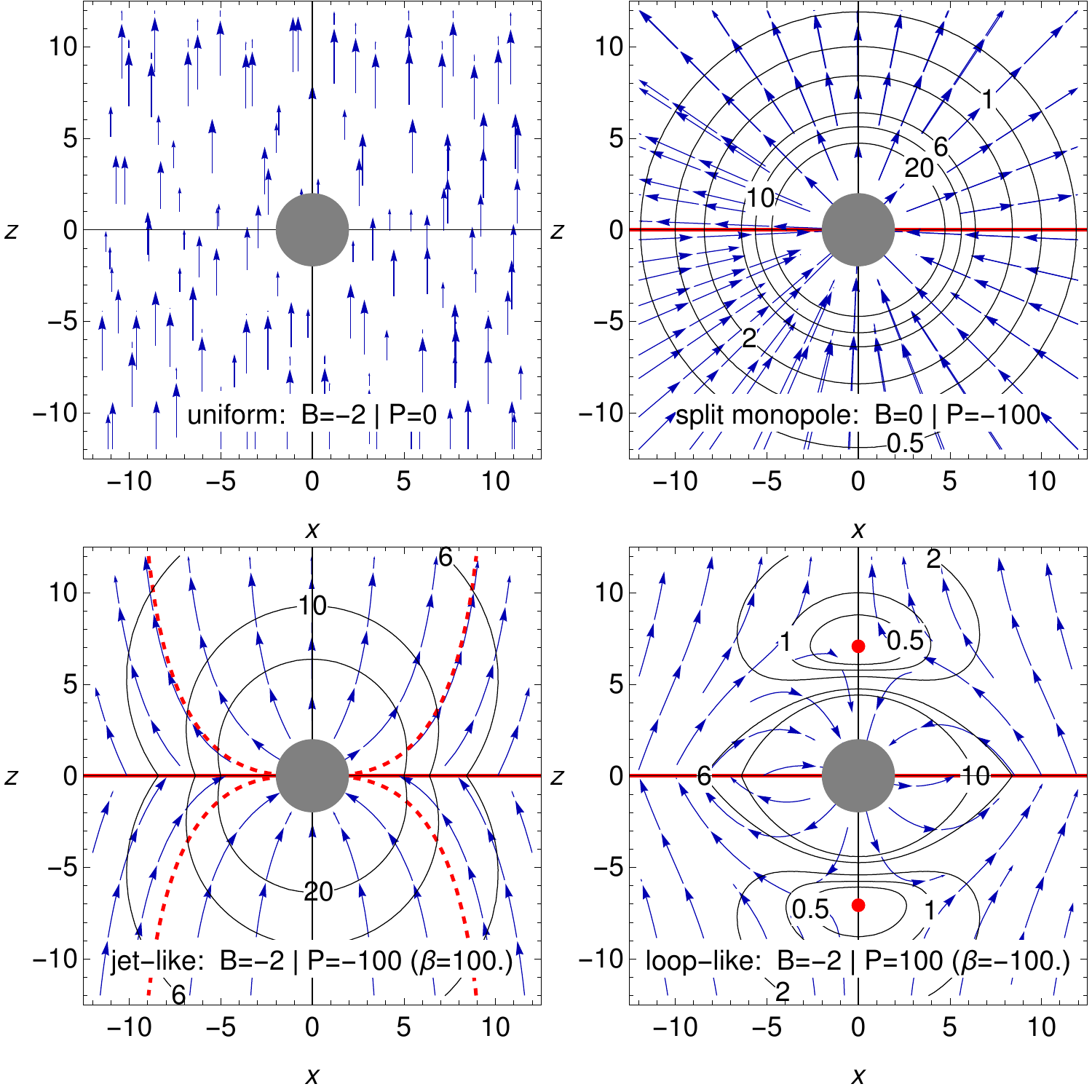}
\caption{Comparison of the magnetic field lines of homogeneous (upper left), Blandford-Znajek split-monopole (upper right), and their superpositions. The magnetic field in the jetlike configuration (bottom left) corresponds to $B P>0$, when the two fields are aligned in the same direction, while the looplike configuration (bottom right) corresponds to $B P<0$, when two fields are aligned in opposite direction. The dark gray circle represents a black hole, the blue arrows indicate the magnetic field direction, and the black solid indicates the contours of the magnetic field magnitude. The BZ split-monopole magnetic field configuration has to be supported by an electric current sheet floating in the $\phi$ direction in the equatorial plane (horizontal red line). Red points located on the vertical axis in the looplike configurations show the positions of the null points, where the magnetic field vanishes. The red dashed lines denote points where the electromagnetic four-potential $A_{\mu}$ turns zero.}
\label{figMF}
\end{figure*}

\section{Combined magnetic field} \label{sec:1}

We consider a black hole of a mass $M$ described by the Schwarzschild metric 
\beq
d s^2 = -f(r) d t^2 + f^{-1}(r) d r^2 + r^2( d \theta^2 + \sin^2\theta d \phi^2), \label{SCHmetric}
\eeq
where $f(r)$ is the lapse function  
\beq 
 f(r) = 1 - \frac{2 M}{r}.
 \label{fun_f}
\eeq
Let the black hole be immersed into an external magnetic field. Without specifying the shape of the field, we assume that the field is stationary and axially symmetric, which allows one to search the four-potential in the form of a linear combination of the timelike and spacelike Killing vectors $\xi^{\mu}$ \cite{Wald:1974:PHYSR4:}
\beq \label{vecpotup}
A^{\mu} = C_1 \xi^{\mu}_{(t)} + C_2 \xi^{\mu}_{(\phi)}, 
\eeq
where $C_1$ and $C_2$ are constants. It should be noted that due to the linearity of Maxwell's equations, a corresponding solution for combined independent field components can be found as a simple superposition of the solutions to individual fields. 
In this paper, we focus on a particular combination of analytic solutions to the magnetic fields, given by the superposition of the uniform and BZ split-monopole magnetic fields, introduced below. 
%

\subsection{Magnetic field components}

The four-vector potential for the uniform magnetic field with an asymptotic value of the strength $B$ and the field lines orthogonal to the black hole's equatorial plane in the Schwarzschild metric has the only non-zero component 
\beq
A_{\phi}^{\rm U} = 
\frac{B}{2} \, r^2 \sin^2 \theta. \label{Auniform}
\eeq
The uniform magnetic field can be interpreted as an external large-scale magnetic field, whose source is located outside and far away from the black hole, i.e. in the region, where the field in a limited local space 
can be seen as approximately homogeneous. 

We combine an external uniform magnetic field with the BZ split-monopole field proposed in \cite{Bla-Zna:1977:MNRAS:}. It is important to note that despite its name and in contrast to the classical magnetic monopole, in the split-monopole case the condition ${\rm div}\, {\vec{B}} = 0$ is not violated, meaning that no magnetic monopole charges are needed.  
For BZ split-monopole magnetic field configuration, the non-vanishing component of the four-vector potential takes the following form
\beq 
A_{\phi}^{\rm SM} = - P |\cos{\theta}|. \label{Asmonopole}
\eeq
Note, that the classical magnetic monopole solution can be recovered by dropping the modulus sign of $\cos{\theta}$ in Eq.~(\ref{Asmonopole}). The difference between the monopole and split-monopole solutions is in opposite directions of magnetic field lines in the southern and northern hemispheres. A BZ split-monopole configuration, which is originally derived from the force-free plasma approximation \cite{Bla-Zna:1977:MNRAS:}, gives an idealized view of magnetic field lines entering into the black hole's horizon through accreting plasma. 

Finally, a combined solution of the two fields, given as a superposition of (\ref{Auniform}) and (\ref{Asmonopole}) can be written in the form 
\bea
A_{\phi}^{\rm U+SM} &=& \frac{B}{2} \, r^2 \sin^2 \theta - P |\cos{\theta}| \nonumber \\ 
&\equiv& \frac{B}{2} \, \left( r^2 \sin^2 \theta - \beta |\cos{\theta}| \right) , \label{Asuperpos}
\eea 
where $\beta = 2P/B$ represents the relative ratio of the strengths of the split-monopole and uniform magnetic fields. %
The sign of $\beta$ governs the relative alignments of the two magnetic field configurations. If we fix $B>0$, then for $\beta>0$, the magnetic field lines of both uniform and split-monopole fields are aligned in the same direction perpendicular to the equatorial plane of the black hole. For $\beta<0$, the magnetic field lines of  uniform and split-monopole fields are aligned in the opposite direction. For $\beta=0$ we recover the uniform magnetic field solution (\ref{Auniform}), while when $|\beta|\gg |B|$, or equivalently $B=0$, we obtain the split-monopole solution (\ref{Asmonopole}). 

One can see the introduced two parametric $B,P$ black hole magnetosphere model (\ref{Asuperpos}) as the most simple expansion of a generic magnetic field around a black hole into spherical harmonic, where only zero (uniform) and first (monopole) components are used. Both uniform and BZ split-monopole components of the combined field $A_{\mu}$ and their combinations satisfy Maxwell's equations and do not violate Gauss's law for magnetism.

\begin{figure*}
\centering
\includegraphics[width=0.45\textwidth]{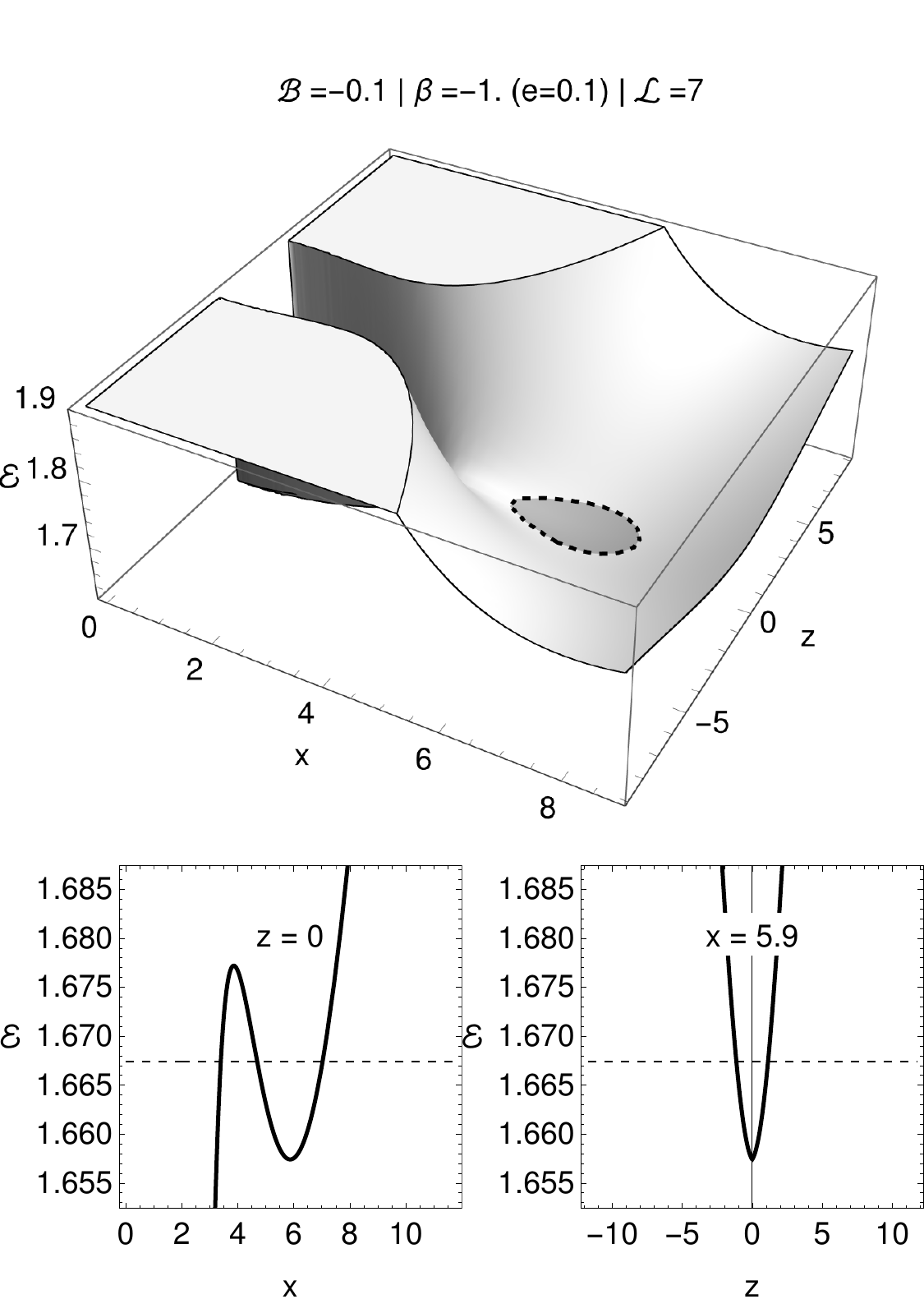}
\includegraphics[width=0.45\textwidth]{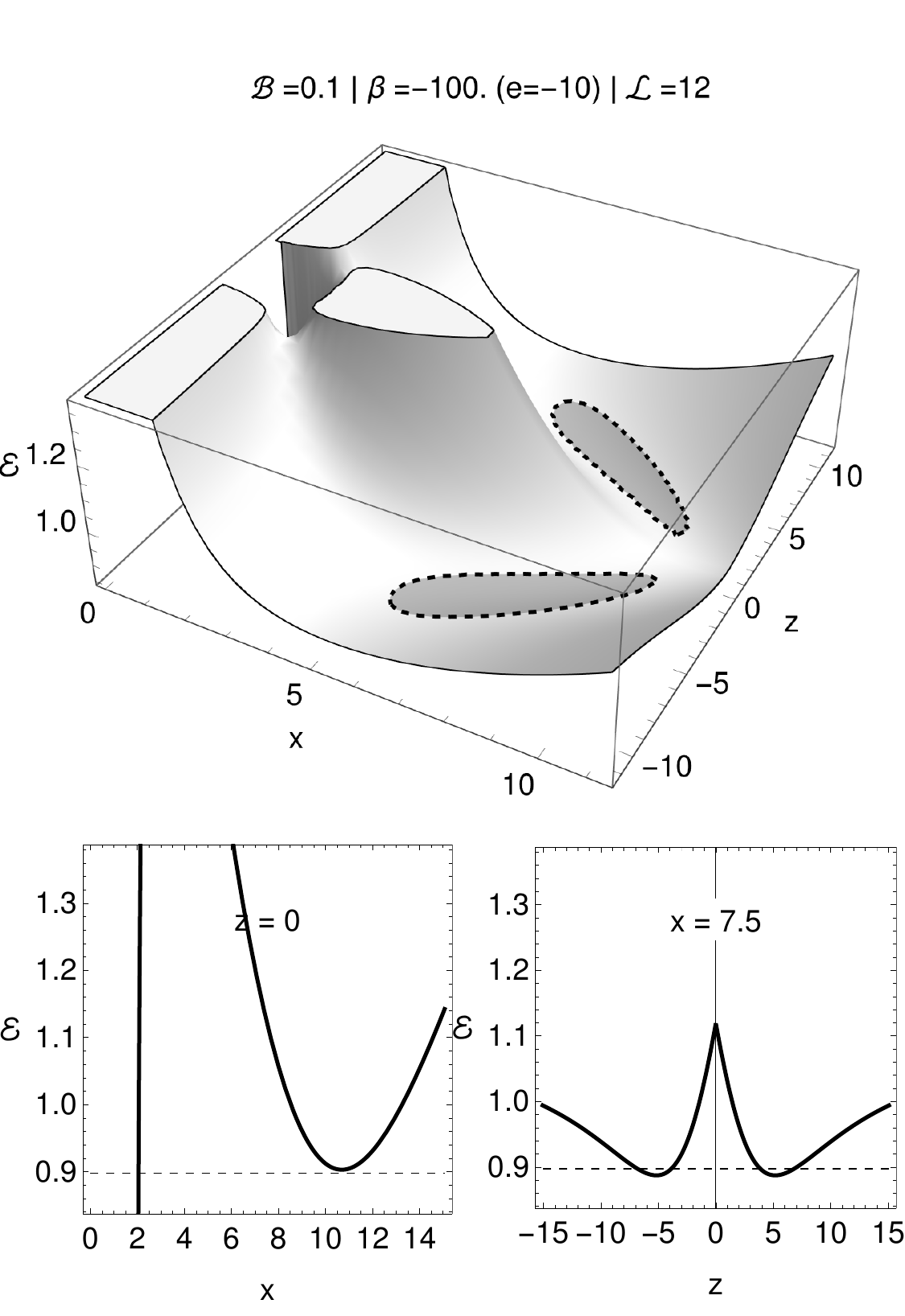}
\caption{An example of the effective potential as a function of $x,z$ for different $\BB$ and $\beta = e/B$ parameters. The left panel for jetlike configuration shows the existence of a single minimum corresponding to a stable orbit at the equatorial plane. The right panel for the looplike configuration shows the existence of the off-equatorial minima with corresponding off-equatorial stable orbits. The effective potential $V_{\rm eff}(x,z)$ will have a "ridge" in the equatorial plane ($z=0$) in a looplike configuration, which will lead to the formation of two minima, one above and one below the equatorial plane.
}
\label{fig_V_eff}
\end{figure*}

The orthonormal components of the magnetic field vector
$ \textbf{B} = (B_{\hat{r}},B_{\hat{\theta}},B_{\hat{\varphi}})$, 
observed by the zero-angular momentum observer (ZAMO) are given by 
\begin{equation}
    B_{\hat{i}}= \eta_{ijk} \sqrt{g^{jj} g^{kk}} F_{jk}, 
\end{equation}
where $ \eta_{ijk} $ is a Levi-Civita tensor in three-dimensional
space. The non-vanishing orthonormal components of magnetic field take the form 
\bea 
B_{\hat{r}} &=& -{B} \left( \cos \theta + \frac{P |\cos \theta |}{r^2 \cos(\theta)} \right), \label{Br} \\
B_{\hat{\theta}} &=& {B} \sin \theta.
\eea
The square of magnetic field strength $\textbf{B}$ as a function of $r$ and $\theta$ for the distant observer takes the form
\beq
\textbf{B}^2 = B^2 +\frac{P^2}{r^4} +\frac{ 2 B P |\cos{\theta}|}{r^2},
\eeq
the first term represents uniform MF, the second -- monopole one, and the third -- its combined contribution.

\subsection{jetlike and looplike configurations}

In Fig.\ref{figMF} we compare the magnetic field lines of the uniform, split-monopole, and the superposition of these two magnetic field configurations with a black hole at the center, as seen by an observer at rest at infinity. The resulting superposed magnetic field configuration significantly differs depending on the relative orientation of the magnetic field components, i.e. the sign of $\beta$.  

In the case with $\beta>0$, that is, when the two field components are aligned, the resulting combination is of a paraboloidal shape. Such a configuration is well motivated by the GRMHD and GRPIC simulations, demonstrating the formation of the paraboloidal geometry of the magnetosphere around a black hole regardless of the shape of an initial magnetic field or accretion configurations (see, e.g. \cite{Kom-McK:2007:MNRAS:,Tch-Nar-McK:2010:APJ:,Kol-Jan:2020:RAG:,Par-Phi-Cer:2019:PRL:}). Moreover, the paraboloidal configuration reminds us of the jet behavior observed in astrophysical black hole systems \cite{Nak-etal:2018:APJ:,Kol-Sha-Tur:2023:EPJC:}. Therefore, the case $\beta>0$ is astrophysically well motivated. Hereafter we call this configuration "jetlike". In Section \ref{sec:Disk} we show that this configuration is optimal for escaping charged particle trajectories. One can see that the field lines close to the black hole in the jetlike configuration extend from the black hole to infinity, which can imply that the energy and angular momentum from the black hole can, in principle, be transferred to infinity along the narrow paraboloidal region. This region also contains a surface of zero electromagnetic potential, as indicated in Fig.\ref{figMF}, which corresponds to the region with a minimal energetic boundary in the effective potential of a charged test particle. One can expect that the jetlike motion along magnetic field lines within this surface is more favorable for charged particles. 

The case with $\beta<0$, where the uniform and BZ split-monopole fields are pointing in opposite directions, forms a configuration with closed field lines connecting the black hole with an equatorial thin disc. 
Such a connection of an event horizon with the accretion disk, if the black hole is rotating, can lead to the energy extraction and subsequent angular momentum transfer from the black hole to the disk \cite{1999ASPC..160..265B,2002A&A...392..469L,2003ApJ...595..109W}. While in the mentioned works the magnetic connection is achieved by the electric currents floating on the inner edge of the disk, here we show that similar field lines can be obtained without currents by a simple combination of the two magnetic fields of opposite directions. However, we leave the discussion of the magnetic connection through a combination of magnetic fields in the Kerr spacetime and subsequent energy extraction to future work. 

Superposition of the fields pointing in opposite directions also leads to the existence of magnetic null points in the black hole vicinity, which play an important role as particle acceleration sites for magnetic reconnection, if a plasma is properly injected \cite{2013IJAA....3...18K,2012CQGra..29c5010K}. We mark these points as red in Fig.\ref{figMF}.  One can find the positions of the magnetic null points taking derivatives from the right-hand side of (\ref{Asuperpos}) with respect to $r$ and $\theta$, giving 
\beq 
B r \sin ^2 \theta =0, \quad \sin 2 \theta  \left(2 r^2 + \frac{\beta}{|\cos \theta |}\right)=0 .
\eeq 
From the above equations, it follows that the magnetic null points can exist only at the poles and be located at the distance $r_{\rm Null}$ from the black hole's center given by
\beq \label{eq:rnull}
r_{\rm Null}^2 = - \frac{\beta}{2}, 
\eeq 
which is valid only in the case when $\beta<0$. Moreover, to have the magnetic null points outside the event horizon, $r_{\rm Null} > r_g$, one should have $|\beta|> 2 r_g^2$, i.e. the BZ split-monopole component of the combined magnetic field  must be sufficiently stronger (by about an order of magnitude) than the external uniform magnetic field component. Such a situation is actually plausible in astrophysical scenarios since the magnetic field generated by the plasma dynamics in the vicinity of a black hole is expected to be stronger than the magnetic field of external origin. Another interesting observation is that the two null points of the considered configuration slightly differ from each other by the directions of ingoing and outgoing magnetic field lines, thus creating an asymmetry between the northern and southern hemispheres. This may cause a difference between the acceleration efficiencies in the magnetic reconnection if plasma injected at the null points leads to asymmetric or even unipolar ejection of relativistic jets, as observed e.g. in \cite{2022MNRAS.517L..86H}.

The realistic values of the parameter $\beta$ depend on the astrophysical system considered. In jetted AGNs a typical magnetic field strength at the event horizon scale of a central supermassive black hole is expected to be around $10^4$G \cite{Daly:APJ:2019:}.. A typical equipartition strength of many spiral galaxies, including our own, is around $10^{-5}$G \cite{2013pss5.book..641B}. However, the galactic fields, which in many cases follow the galactic spiral arms, are amplified toward the center \cite{2013pss5.book..641B,2021ApJ...923..150L}. Thus, if the galactic magnetic field is the only source of the external field, the ratio of the internal to  external magnetic field components can be rather large, reaching $\beta = 10^7-10^9$. In the case of the Galactic center supermassive black hole SgrA* with a magnetic field around $10$G \cite{Eckart-etal:2017:FOP:}, one gets the parameter $\beta_{\rm SgrA*} = 10^4 - 10^6$. However, the situation can change significantly for binary black hole systems. In a certain separation distance between the two components of the binary, the primary black hole can be immersed into the magnetic field of the companion object. If the companion object is a neutron star, as in the case of black hole microquasars, the external field may start to dominate the internal field of the black hole accretion disk, leading to $\beta < 1$. However, such a situation is likely to be observed at the final stages of the binary evolution. Thus, the parameter $\beta$ can have a wide range of values from $\sim 1$ to $10^9$. 

It is important to note, that for larger $\beta$, e.g. $\beta \sim 10^9$, the shape of the field lines is essentially the same, as e.g. in Fig.~\ref{figMF} shown for $\beta \sim 10^2$. The whole picture for larger $\beta$ will just be rescaled to larger distances, given by the factor of $\sqrt{\beta/2}$, as now the transition from the split-monopole to the uniform field will appear at larger radii, but the configurations of the combined fields will not change. For example, according to Eq.~\ref{eq:rnull} and Fig.~\ref{figMF}, the null-points located at the poles at $r_{\rm Null} \approx 7.1$ for $\beta = 100$; essentially the same shapes of the field lines can be obtained for $\beta = 10^9$ with $r_{\rm Null} \approx 2.2 \times 10^4$, where the axes are rescaled by the factor of $\sqrt{\beta/2} \sim 3.1 \times 10^4$. 
Therefore, the results presented in the article for particular values of $\beta > 1$ are equally valid also for larger values of the parameter $\beta$.

One can see that both field configurations obtained from the combination of uniform and BZ split-monopole magnetic fields can be astrophysically relevant in various scenarios involving black holes. Regarding AGNs with supermassive black holes at their centers, while the jetlike configuration can apparently be related to the jet formation scenarios of radio-loud AGNs, the looplike configuration can be related to radio-quiet AGNs,  where energy release occurs without apparent jets \cite{2019NatAs...3..387P}. The possible connection of the two AGN types and their regimes of energy emission with the role of an external magnetic field component in relation to the internal magnetic field of the source needs further investigation, which is beyond the scope of this paper.

\section{Dynamics of charged particles} 
\label{sec:2}

\begin{figure*}
\centering
\includegraphics[width=\textwidth]{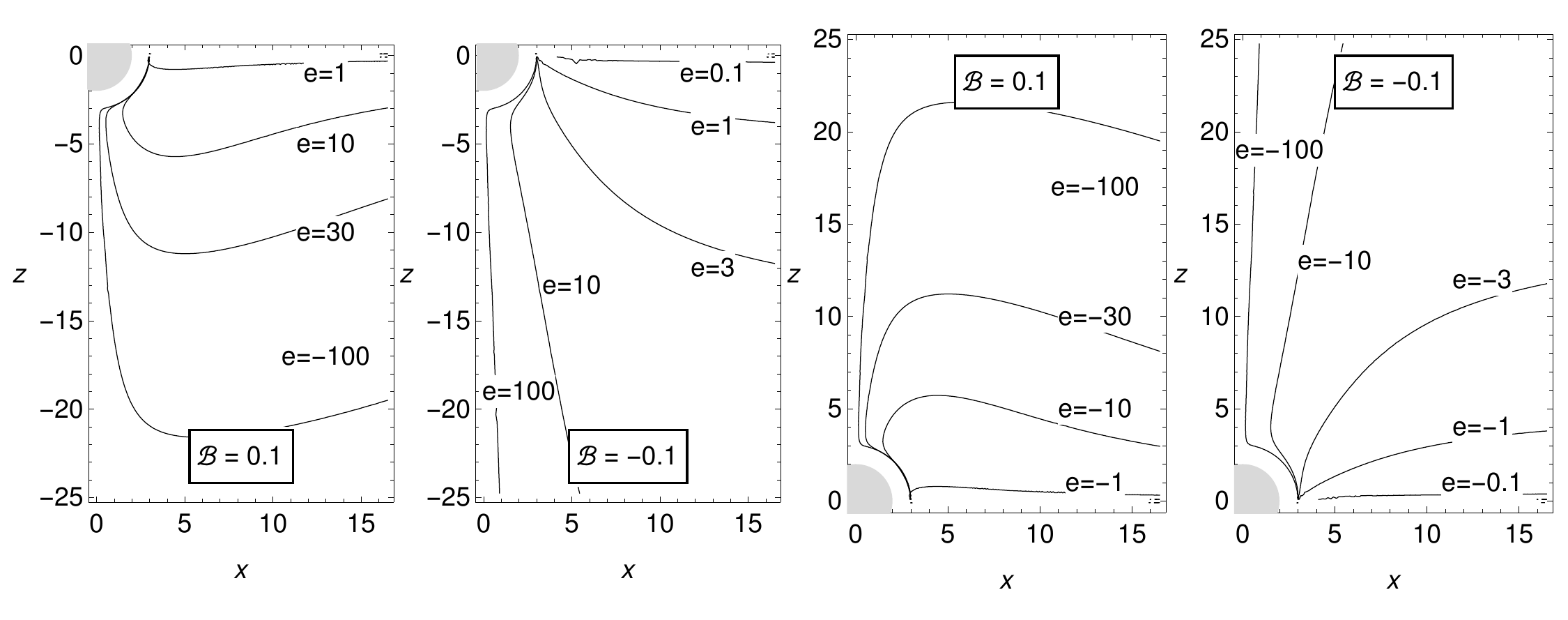}
\caption{Positions of stationary points of the off-equatorial effective potential $V_{\rm eff}(x,z)$ in $x$-$z$ plane for various values of the magnetic field parameters. %
One can see that the off-equatorial circular orbits can exist in both jetlike and looplike configurations. With increasing the split-monopole parameter $e$, the off-equatorial stable orbit shifts further away from the black hole. 
In the jetlike configuration ($\cb e >0$) the change of the off-equatorial orbit is monotonic, while in the looplike configuration ($\cb e <0$) there always exists a maximum distance from the equatorial plane. Due to the symmetries of the effective potential, discussed after (\ref{VeffCharged}), in the chosen $\cl>0$ case, the circular orbits exist above the equatorial plane for $e<0$ while below the plane for $e>0$. There are no circular off-equatorial orbits below the photon orbit $r<3$.}
\label{fig:offmin}
\end{figure*}
\begin{figure*}
\includegraphics[width=\textwidth]{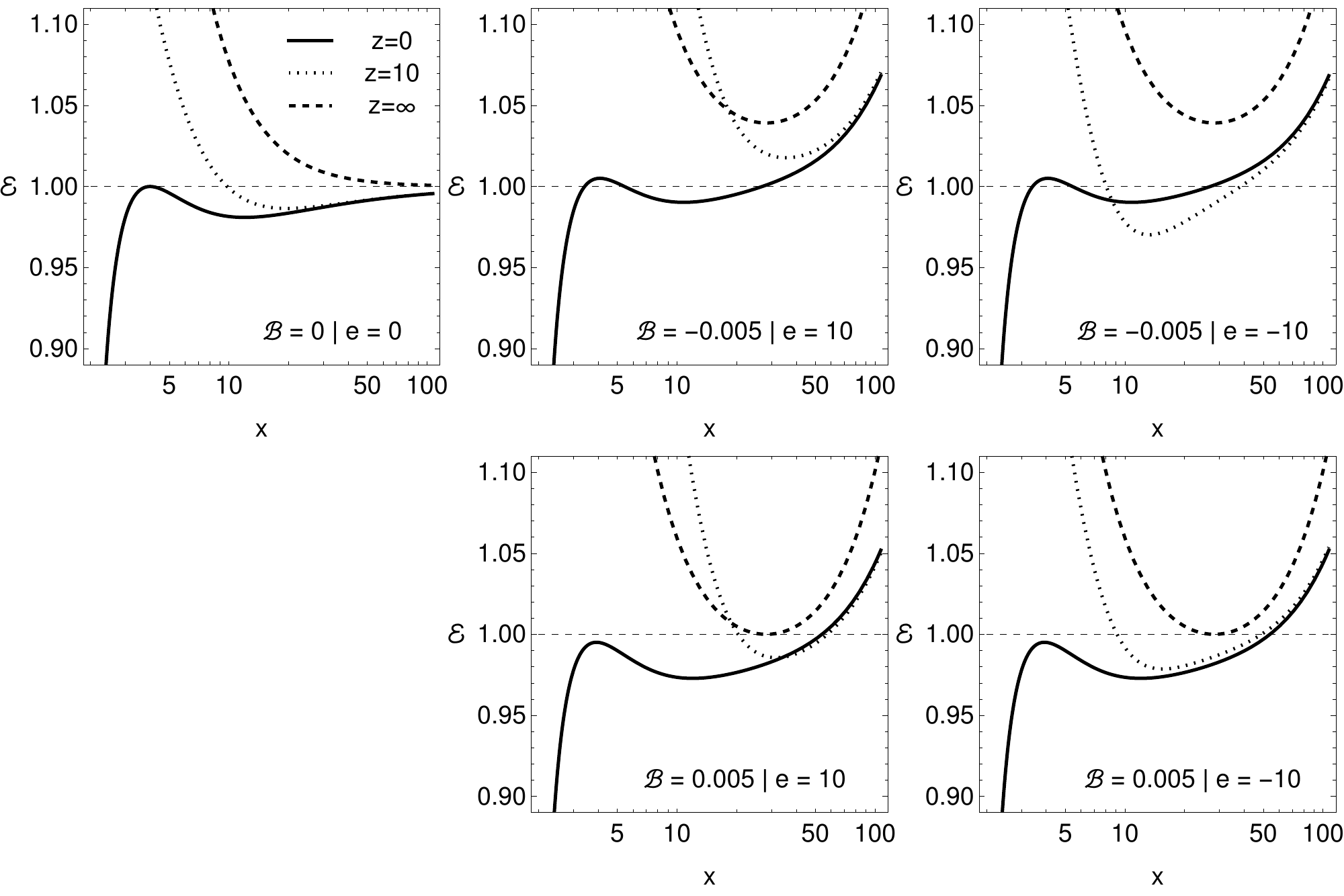}
\caption{The horizontal $x$ cross section of the effective potential $V_{\rm eff}(x,z)$ for different vertical values $z$. A particle can escape to infinity if its energy satisfies the condition (\ref{escape}).}
\label{fig:escape}
\end{figure*}

\subsection{Equations of motion}

In this section, we explore the dynamics of a charged test particle in combined uniform and BZ split-monopole magnetic fields around a Schwarzschild black hole. The equation of motion of a particle with a charge $q$ and mass $m$ is given by the Lorentz equation
\beq 
m\frac{ D u^\mu }{\d \tau} = q F^{\mu}_{\ \nu} u^{\nu}, 
\eeq
where $D$ denotes a covariant derivative, $u^{\mu}$ is a four-velocity of a test particle satisfying the condition $u^{\mu} u_{\mu} = -1$, $\tau$ is the proper time of the particle, and $F_{\mu \nu} = A_{\nu,\mu} - A_{\mu,\nu}$ is the antisymmetric Faraday tensor of electromagnetic field.

The Lagrangian density for the charged particle motion is given by 
\begin{equation}
    {L} = \frac{1}{2} g_{\mu\nu} \frac{d x^{\mu}}{d\tau} \frac{d x^{\nu}}{d\tau} + \frac{q}{m} A_{\mu} \frac{d x^{\mu}}{d\tau}\, .
    \label{eq_lagran}
\end{equation}
Since (\ref{eq_lagran}) does not depend explicitly on $t$ and $\varphi$ variables, these two symmetries give rise to two integrals of motion, namely the total energy and the canonical angular momentum, respectively 
\bea
 \EE &=& f(r) \, \frac{d t}{\d \tau}, \\
 \LL &=&  r^2 \sin^2\theta \, \frac{d \phi}{d \tau} + 
 \BB \, \left( r^2 \sin^2\theta - \beta | \cos\theta | \right). \label{angmom}
\eea
Here we use specific energy $\EE$, specific axial angular momentum $\LL$ 
\beq
\ce = \frac{E}{m}, \quad \cl = \frac{L}{m},
\label{EL}
\eeq
and and magnetic field  parameters $\BB,\beta,e$,
\beq
\cb = \frac{q B}{2m}, \quad e = \frac{q P}{m}, \quad \beta = \frac{e}{\cb} \label{Beb}
\eeq
which reflects a relative relationship between Lorentz and gravitational forces, see discussion of Eq.~(\ref{BB-param})

\subsection{Effective potential} \label{sec:eff}
Using the normalization of the four-velocity, one can find the effective potential in the form
\bea 
&&V_{\rm eff} \equiv V_{\rm eff} (r,\theta) = \nonumber \\ 
&&  f(r) \Bigg[1+ \frac{ \bigg(\LL- {\cal B}\, \left( r^2 \sin^2 \theta - \beta |\cos{\theta}| \right) \bigg) ^2}{r^2 \sin^2 \theta}\Bigg]. \label{VeffCharged}
\eea 
In general, the Lorentz force acting on the charged particle moving around black hole in external magnetic field can be either attractive or repulsive with respect to the gravitational force \cite{Tur-Stu-Kol:2016:PHYSR4:}. Since a combination of magnetic field components gives two different configurations, in total one can distinguish four qualitatively different types of motion. 
If $\beta < r^2 \sin^2\theta / |\cos{\theta}|$ the attractive Lorentz force (i.e. directed towards the black hole) is obtained for $\LL>0, \BB<0$, which is also equivalent to $\LL<0, \BB>0$ due to above-mentioned symmetry. Inversely, the repulsive Lorentz force (i.e. directed outward from the black hole) is obtained for $\cl>0, \cb>0$, or equivalently for $\cl<0, \cb<0$. 
On the other hand, if $\beta > r^2 \sin^2\theta / |\cos{\theta}|$, we have an inverse situation, so that the attractive Lorentz force is obtained for $\LL>0, \BB>0$, or equivalently to $\LL<0, \BB<0$ and the repulsive Lorentz force is obtained for $\cl>0, \cb<0$, or equivalently for $\cl<0, \cb>0$.  
In general, one can distinguish the following three situations:
\begin{itemize}
\item[--] {\it attractive Lorentz force}, $\BB r^2 \sin^2 \theta < \BB \beta |\cos\theta|$.
\item[+] {\it repulsive Lorentz force},  $\BB r^2 \sin^2 \theta > \BB \beta |\cos\theta|$.
\item[0] {\it null configuration}, $r^2 \sin^2 \theta  = \beta |\cos\theta|$.
\end{itemize}
Since $r$ and $\theta$ are dynamical parameters, while $\BB$ and $\beta$ are fixed, in a conservative system (with $\EE$ and $\LL$ conserved), it is possible to find the trajectory, along which the Lorentz force changes direction.  This is a distinguishable feature of the combined magnetic field that cannot be observed in the case of individual field configurations \cite{Kol-Stu-Tur:2015:CLAQG:,Kol-Bar-Jur:2019:RAGtime:}.
Due to such symmetries of the effective potential and without loss of generality, hereafter, we fix $\cl>0$, while the magnetic parameters $\BB$ and $\beta$ can be positive or negative. 

The effective potential (\ref{VeffCharged}) shares the background symmetries of the spacetime metric and those of the superposition magnetic field. 
It is independent of the coordinate $\phi$ and diverges at the event horizon $r = 2 M$. We exclude the regions of the horizon and divergent points from our considerations; the effective potential is positive everywhere outside the horizon. Due to background symmetries, we focus on the region $\theta\in(0,\pi/2)$, which allows us to remove the modulus sign appearing due to the BZ split-monopole solution in the considered region. In Fig.\ref{fig_V_eff} we plot the effective potential as a function of $r$ for the positive and negative values of the magnetic field. We plot the effective potential as a function of Cartesian coordinates $x$-$z$ in Fig.~\ref{fig_V_eff}. For that, we use the coordinate conversion 
\beq 
x = r \, \cos{\phi} \, \sin{\theta}, \,\,\, y = r \,\sin{\phi} \, \sin{\theta}, \,\,\, z = r \, \cos{\theta},
\eeq 
which is a standard transformation. 

The effective potential represents an energetic boundary given by 
\beq
 \EE^2 = V_{\rm eff} (r,\theta). \label{MotLim}
\eeq
The stationary points of $V_{\rm eff}(r,\theta)$, defining maxima or minima, are given by the derivatives 
\beq
  \partial_r V_{\rm eff}(r,\theta) = 0, \quad \partial_\theta V_{\rm eff}(r,\theta) = 0.  \label{extrem}
\eeq
These equations read explicitly in the following form
\bea 
&& M r^2 - 2 F  \BB r^2 (r - 2 M) - \frac{F^2 (r - 3 M)}{\sin^2 \theta} = 0, \label{drVeff}\\
&& F \bigg[ \left( r^2 \BB + \frac{\LL + \beta \BB |\cos\theta|}{\sin^2\theta} \right) \cos \theta  + \frac{\beta  \BB}{|\cos\theta|} \bigg] = 0,  \label{dthVeff}
\eea
where 
\beq 
F = \mathcal{L}-\mathcal{B} \left(r^2 \sin ^2\theta -\beta |\cos \theta |\right).
\eeq 
In contrast to the homogeneous magnetic field case \cite{Kol-Stu-Tur:2015:CLAQG:}, in the combined magnetic field the effective potential has minima also outside of the equatorial plane. 
Further, we study the orbital motion of charged particles, including off-equatorial conditions.

\begin{figure*}
\includegraphics[width=1\textwidth]{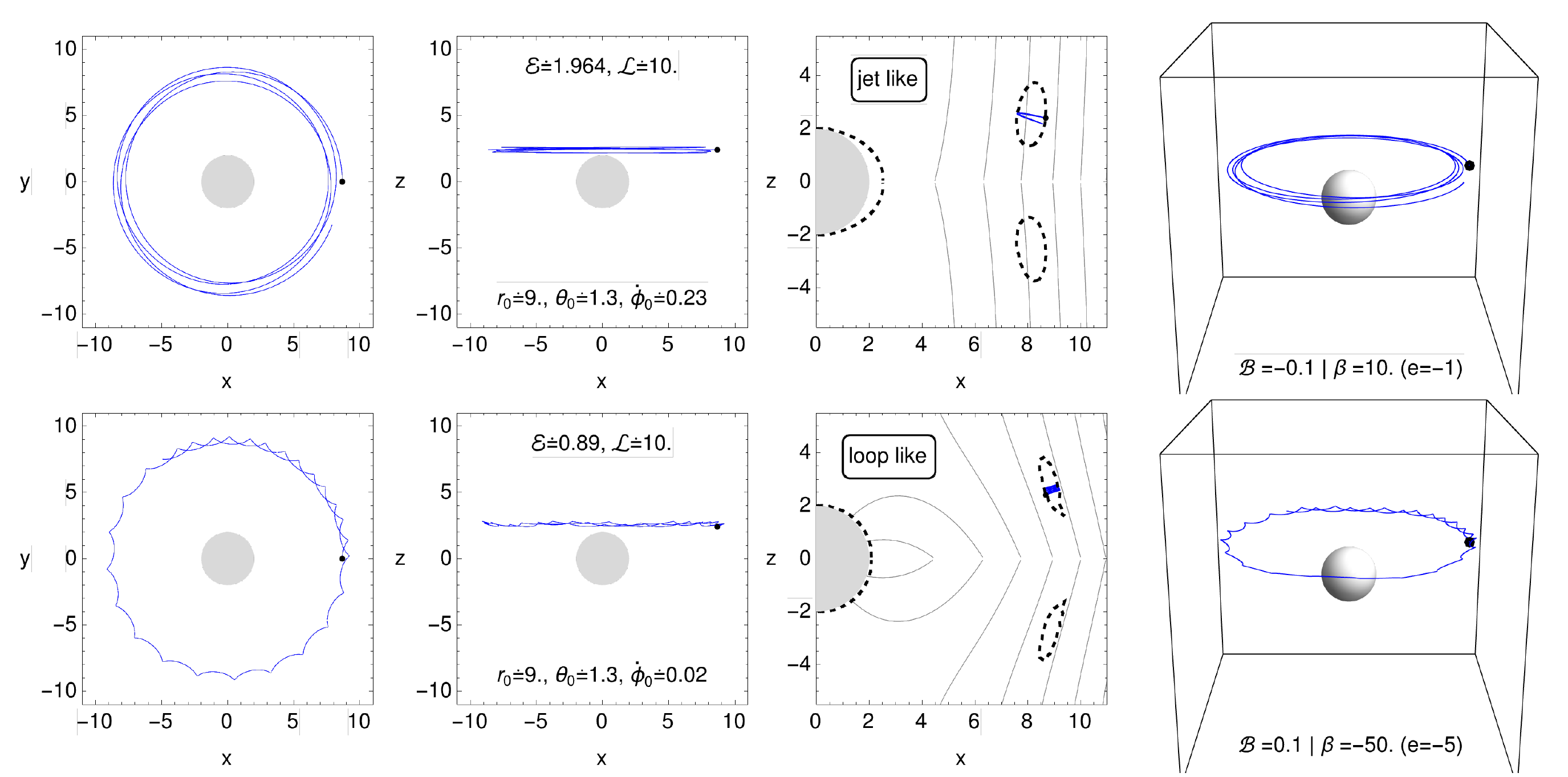}
\caption{Examples of two perturbed off-equatorial stable orbital trajectories (blue lines) around Schwarzschild black hole (gray circle) in the jetlike (first row) and looplike (second row) magnetic field configurations (thin gray lines). The energetic boundary of the particle is shown by the dashed black lines. Due to the presence of an attractive Lorentz force in the jetlike configuration, the orbital motion in the $\phi$-direction is much faster than in the looplike configurations, where one can even see retrograde curls due to the existence of a repulsive Lorentz force. %
The positions of the off-equatorial orbits are plotted in Fig.~\ref{fig:offmin}.}
\label{fig:offtraj}
\end{figure*}

\subsection{Equatorial and off-equatorial orbital motion}

Equatorial motion with $\theta = \pi /2$ in a combined magnetic field (\ref{Asuperpos})  reduces to the motion in the uniform magnetic field (\ref{Auniform}), due to the vanishing of the BZ split-monopole field component at the equatorial plane. Such a motion has been studied in detail in the past (see, e.g. \cite{Kol-Stu-Tur:2015:CLAQG:}). 

Due to the presence of a homogeneous component in the superposition field, the motion of the charged particle is always bounded in the radial direction around the equatorial plane $\theta = \pi / 2$, due to the term $\BB^2 r^2$ appearing in the limit of $r\rightarrow\infty$ in the equatorial region \cite{Kol-Stu-Tur:2015:CLAQG:}.  Therefore, charged particles in the superposition magnetic field can escape from the black hole only in vertical directions, i.e. along the magnetic field lines, as expected. 

On the other hand, the effective potential has more than one extremum in $\theta$, given by (\ref{dthVeff}), which implies that stable circular orbits can exist in both the equatorial and off-equatorial planes. From (\ref{dthVeff}) it follows that $\cos\theta=0$, leading to the equatorial orbits with $\theta = \pi/2$. Another solution of (\ref{dthVeff}) can be obtained by equalizing parentheses to zero, which leads to the cubic equation in $|\cos\theta|$ 
\beq 
(|\cos\theta_{\rm off}|^3 - |\cos\theta_{\rm off}|) r^2 \BB - |\cos\theta_{\rm off}| \LL - \BB \beta = 0.
\eeq 
The above equation has the only real root, that is 
\beq  \label{costhetafull}
|\cos\theta_{\rm off}| = \frac{H^{2/3}/r^2 \BB + 12^{1/3} (r^2 \BB + \LL)}{18^{1/3} J^{1/3}}, 
\eeq 
where 
\beq 
H = 9 r^4 \BB^3 \beta + \sqrt{3 r^6\BB^3 (27 r^2 \BB^3 \beta^2 - 4 (r^2 \BB + \LL)^3)}.
\eeq
The analytical solution of obtained equation is non-trivial, however at the regions close to the equatorial plane, one can find the following approximate solution for the off-equatorial stable orbits 
\beq 
|\cos\theta_{\rm off}| \approx - \frac{\beta \BB}{r^2 \BB + \LL}.
\eeq 
Since the right-hand side of this equation must be positive, it puts restrictions on the signs of $\BB$ and $\beta$ for given $\LL$. Obviously, the existence of the off-equatorial orbits is due to the contribution of the BZ split-monopole magnetic field \cite{Kol-Bar-Jur:2019:RAGtime:}, since such orbits cannot exist in the pure uniform magnetic field case \cite{Kol-Stu-Tur:2015:CLAQG:}. 
In a pure BZ split-monopole case (without uniform magnetic field component), the condition for the existence of the off-equatorial stable orbits is $\BB\beta<0$, while for $\BB\beta>0$  stable orbits can exist only at the equatorial plane. In a superposition magnetic field, the situation is slightly different. For $\LL>0$ (we are free to choose the sign of angular momentum, due to the spherical symmetry of the black hole), to hold $|\cos\theta_{\rm off}|>0$, one may have the following three conditions
\begin{itemize}
    \item[(i)] $\beta<0$, \,\,\, $\BB>0$; 
    \item[(ii)] $\beta>0$, \,\,\, $\BB<0$, \,\,\, $\LL > |r^2 \BB|$;
    \item[(iii)] $\beta<0$, \,\,\, $\BB<0$, \,\,\, $0<\LL<|r^2 \BB|$.
\end{itemize}
Since the sign of the magnetic parameter $\BB$ depends on both the sign of the particle's charge and magnetic field direction, while $\beta$ is independent from the sign of the charge, one may have all three situations in either combination of the two field lines. 

Angular momentum at a circular orbit is then given by 
\bea 
&& \LL_{\rm c.o.} = -\mathcal{B} \left(\beta |\cos\theta| + \frac{M r^2 \sin ^2\theta }{r-3 M}\right) \nonumber \\ 
&& \pm \frac{r \sin \theta  \sqrt{r^2 \mathcal{B}^2 \sin ^2\theta (r-2 M)^2+M (r-3 M)}}{r-3 M},  \label{LLco}
\eea 
corresponding to both equatorial and off-equatorial orbits, in which $|\cos\theta|$ is either zero (for equatorial orbits), or given by (\ref{costhetafull}) (for off-equatorial ones). 

In Figure \ref{fig:offmin}, we show the locations of stable circular orbits around a black hole in a combined field for various magnetic parameters. Both jetlike and looplike configurations have off-equatorial plane minima.

The motion of charged particles in a combined magnetic field is always bounded in the $r$ direction, due to the ~$\cb^2r^2$ term present in the effective potential (\ref{VeffCharged}). However, the motion can be open in the vertical $z$ direction if allowed by the energetic boundary condition (\ref{MotLim}). Thus the charged particles can escape to infinity along the $z$-axis if the particle's energy is large enough
\beq
 \ce \geq \ce_{\rm flat(min)} = 
\Big\{ 
\begin{array}{l @{\quad} c @{\quad} l} 
1 & \textrm{for} & \cb \geq 0, \\ 
\sqrt{1 + 4 \cb \cl} & \textrm{for} & \cb < 0, \\ 
\end{array} \label{escape}
\eeq
where $\ce_{\rm flat(min)}$ defines the minimum energy of the charged particle at infinity; see Fig.~\ref{fig:escape}. At the equatorial plane, $z=0$ and at infinity $z\rightarrow\infty$ the BZ split-monopole  component vanishes, making the energetic escape condition in the combined field the same as in the case of the uniform magnetic field \cite{Kol-Stu-Tur:2015:CLAQG:}. 
In Fig.\ref{fig:offtraj} we show examples of the off-equatorial trajectories of charged particles in jetlike and looplike configurations. 

\begin{figure}
\centering
\includegraphics[width=0.35\textwidth]{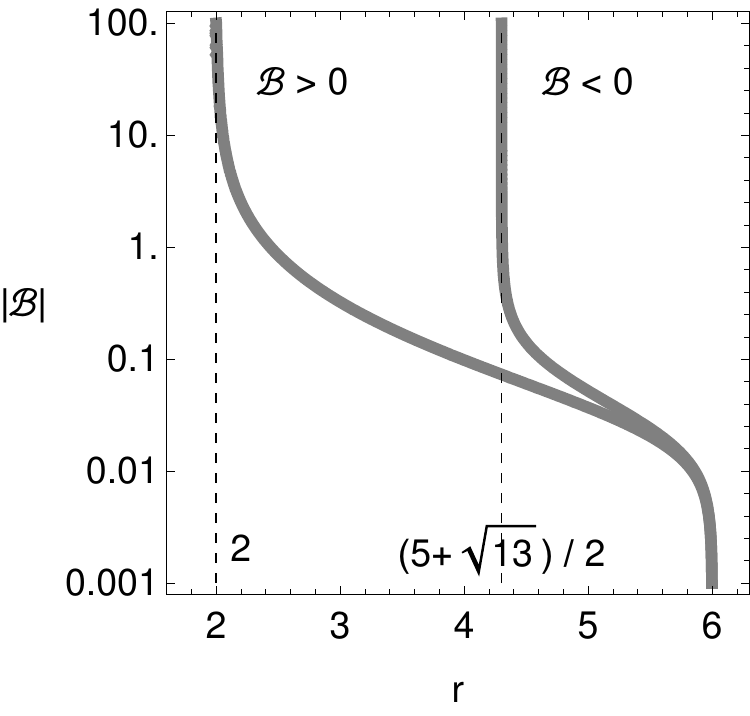}
\caption{Radial position of the effective potential minima at the equatorial plane as a function of uniform magnetic field parameter $\cb$ \cite{Kol-Stu-Tur:2015:CLAQG:}. The BZ split-monopole component of the magnetic field $\beta$ ($e$) vanishes at the equatorial plane and has no influence on the radial stability of the circular equatorial orbit, although having a strong effect on the vertical stability as can be seen from Fig.~\ref{fig_V_eff}.}
\label{fig:inmin}
\end{figure}

\begin{figure*}
\includegraphics[width=1\textwidth]{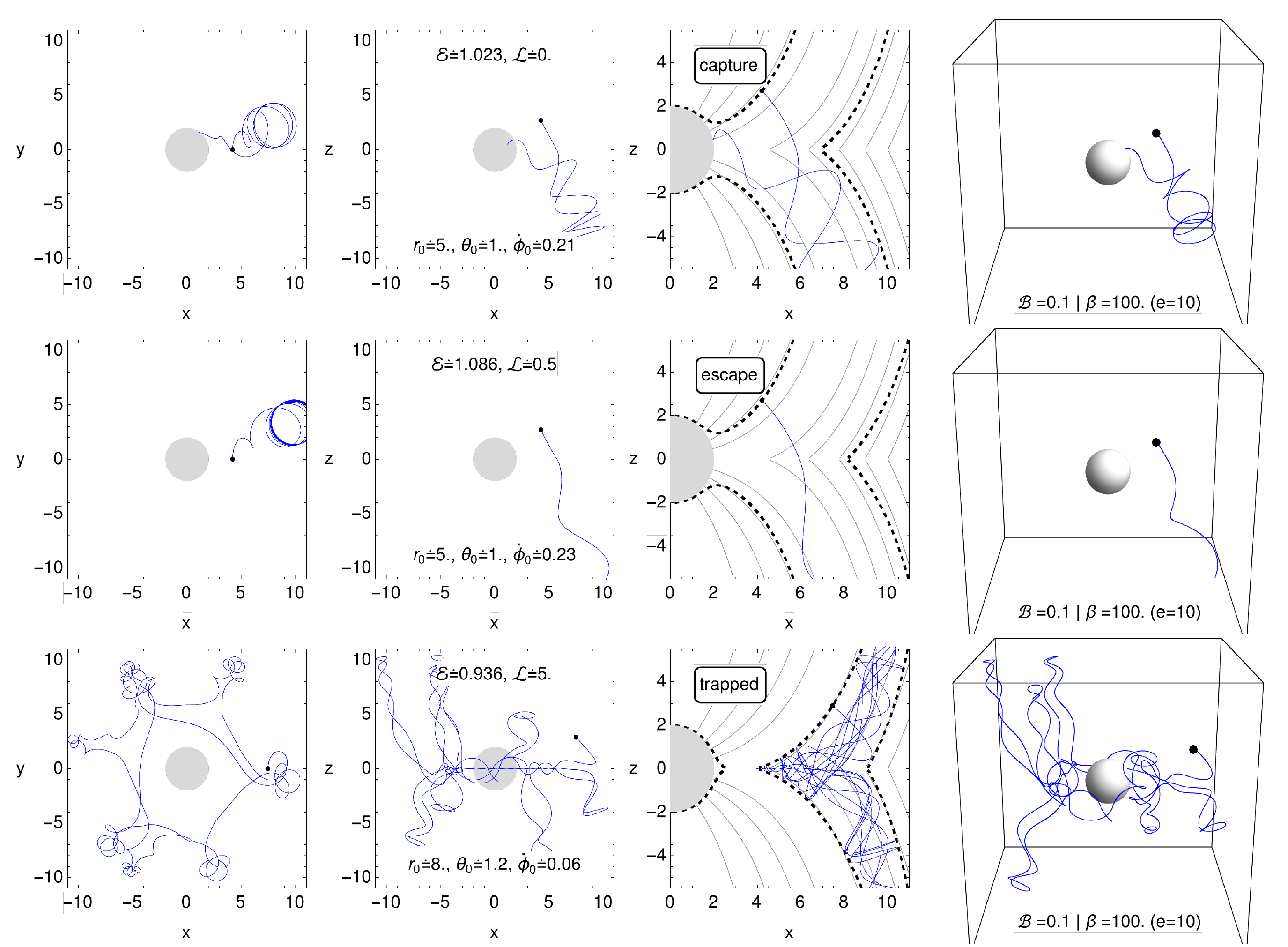}
\caption{
Examples of charged particle trajectories around black hole in a combined magnetic field of the jetlike configuration corresponding to the capture by black hole (first row), escape to infinity along the magnetic field lines (second row), and the trapped oscillatory state (third row).}
\label{fig_sp_motion}
\end{figure*}
\begin{figure*}
\includegraphics[width=1\textwidth]{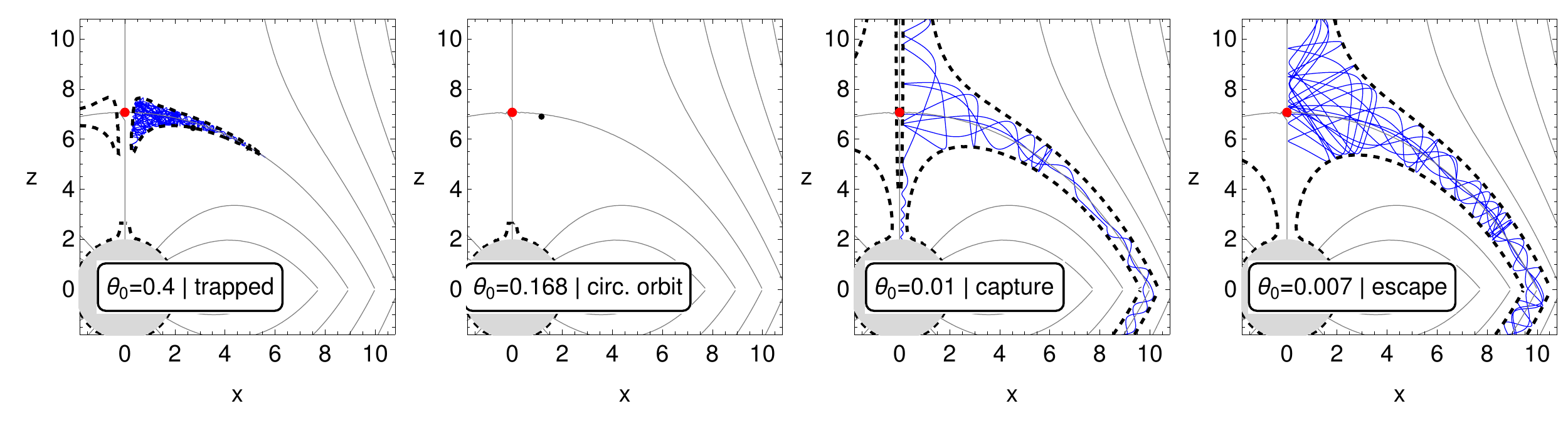}
\includegraphics[width=1\textwidth]{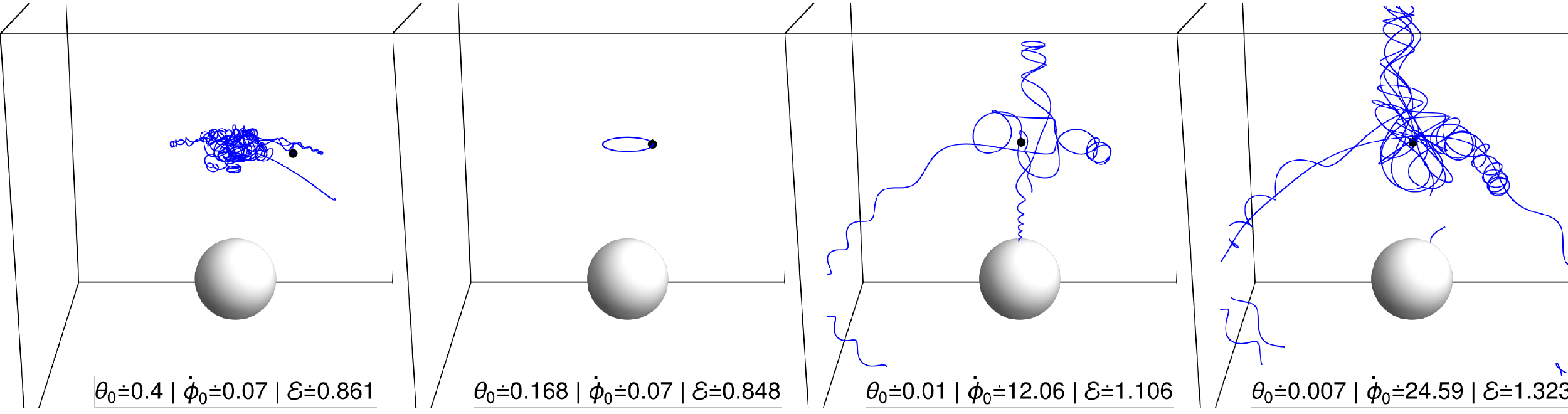}
\caption{Charged particle motion around magnetic null points. We chose the magnetic parameters for the looplike configuration to be $\cb=1$, $e=-100$ $(\beta=e/\cb=-100)$, hence the magnetic null point is located at $r_{\rm null} = 5\sqrt{2} \doteq 7.07$. For the four presented trajectories, we keep the same angular momentum $\cl=100.059$ and initial position $r_0=7$ at different inclinations of $\theta_0$. Keeping the same $\cl$ means that the shape of the effective potential $V_{\rm eff}(r,\theta;\cl)$ is the same in all cases, and a change in $\theta_0$ will only elevate the energy of the charged particle revealing different energy boundaries for particle's trajectory from eq.~(\ref{MotLim}). Then the off-equatorial minimum is located at $r=7,\theta\doteq0.168$. In the first column, we show the particle oscillating around the off-equatorial minimum, in the second column the particle is at the circular off-equatorial orbit, in the third column the particle is captured by the black hole, and in the fourth column, it escapes to infinity along the vertical $z$ axis. The magnetic null point is not a barrier for charged particle dynamics, as the charged particles move along the magnetic field lines. 
}
\label{fig:null-traj}
\end{figure*}

\subsection{Innermost stable circular orbits and generic trajectories} 

One of the most important characteristics of particle motion in general relativity is the presence of the innermost stable circular orbit (ISCO), below which all orbits are unstable. In a pure Schwarzschild black hole case (without magnetic field), the ISCO is located at $r=6M$. When electromagnetic interaction is taken into account, the ISCO may shift towards or outward from the black hole depending on the chosen magnetic field configuration and the direction of the Lorentz force \cite{Tur-Stu-Kol:2016:PHYSR4:,Kov-etal:2010:CLAQG:}. The ISCO radius at the equatorial plane, $r_{\rm \frac{\pi}{2} ISCO}$ is located closer to the black hole than the corresponding last stable off-equatorial orbit $r_{\rm offISCO}$. Since in the equatorial plane, the BZ split-monopole component of the magnetic field vanishes, the ISCO is actually given by the relation corresponding to the uniform magnetic field case \cite{Kol-Stu-Tur:2015:CLAQG:}. 
The ISCO in a general plane can be found by using the second derivative of the effective potential 
\bea 
\frac{d^2 V_{\rm eff}}{dr^2} &=& 0 = {\cal B} ^2 \Bigg[ \frac{2 \beta \big| \cos \theta \big|}{r^3} \bigg(4+ \frac{3 \beta \left(r-4 \right) \big| \cos \theta \big| }{r^2 \sin ^2 \theta} \bigg) \nonumber 
\\  
&+&2 \sin^2 \theta \Bigg] + \frac{4 {\cal B}  \beta \cl}{r^3} \bigg[ \frac{3 \beta \left( r-4\right) \big| \cos \theta \big|}{r^2 \sin^2 \theta} +2 \bigg] 
\nonumber \\  
&+& \frac{2}{r^3} \bigg[ \frac{3\cl^2 \left(r-4 \right)}{r^2 \sin^2 \theta} -2 \bigg].
\eea
Substituting here $\theta = \pi/2$ and $\LL = \LL_{\rm c.o.}$ from (\ref{LLco}) one gets the equation for the ISCO radius 
\bea 
&& 6-r- 2\BB^2 r  (2 r^3 - 9 r^2 + 8 r - 12) \nonumber 
\\ 
&& +2\BB (r-6) \sqrt{\BB^2 r^2 (5 r^2 - 4r + 4) + 2 r} = 0.
\eea
Solving the above equation with respect to $r$ gives the ISCO radius at the equatorial plane, which as expected, coincides with the results of \cite{Kol-Stu-Tur:2015:CLAQG:}. 
We plot the dependence of the ISCO on positive and negative $\cb$, corresponding to the repulsive and attractive Lorentz force in Fig.\ref{fig:inmin}.

In general, the fate of a charged particle is defined by the initial conditions and the corresponding effective potential, which allows one to separate bounded, escaping, and collapsing orbits in parameter space. In Fig.\ref{fig_sp_motion} we show some interesting examples of collapsing (first row), escaping (second row), and bounded (third row) trajectories in a combined magnetic field of jetlike configuration. Similar fates of particles can obviously be observed in the looplike configuration, too. Since the looplike configuration can have magnetic null points, it is interesting to study the behavior of charged particles in the vicinity of such points. In Fig.\ref{fig:null-traj} we plot the example of charged particle trajectories around the null points in the looplike configuration. However, in the Schwarzschild case, no special behavior of the particle is observed around the magnetic null points. 

From an analysis of trajectories, one can extract important information on the role of magnetic field configuration on the fate of the particle. In the next section, we will show that the jetlike combination of magnetic fields supports escape trajectories of charged particles, while the looplike configuration more likely supports accretion into the Schwarzschild black hole.

\section{Ionized Keplerian disks \label{sec:Disk}}

In this section, we consider the ionization of particles from Keplerian accretion disk orbiting a Schwarzschild black hole in the presence of a combined magnetic field. 
In realistic astrophysical situations, a black hole accretion disk is expected to be fully ionized due to the very high temperature and density of the accretion flow, satisfying the global neutrality condition. Due to the quasineutrality of the plasma, the angular momentum of an accretion disk is governed by neutral circular geodesics. However, at the edges of an accretion disk, one can always expect some individual particles to be separated from the main accretion flow. Such a charge separation can be caused by various factors and is not statistically excluded. 
In these cases, the charged particles, initially following neutral geodesics within the accretion flow, after separation could start feeling the electromagnetic forces and be governed by the Lorentz equation. Depending on the strength of the electromagnetic interaction and initial conditions, the charged particles can form low-density bounded structures above the disk (similar to Earth’s radiative belts) or escape to infinity if their energies are large enough. 

Moreover, in a hot and dense accretion plasma, the processes of nucleosynthesis can occur \cite{2008ApJ...681...96H}, which include the production of neutrinos, electron-positron pairs, neutrons, and heavier nuclei. For example, free neutrons can be produced in the accretion torus in processes $p + e^{-} \rightarrow n + \nu_e$ and $p + \Tilde{\nu}_e  \rightarrow n + e^{+}$ \cite{2017Galax...5...15J}, for the temperature and density being $\sim 10^{ 11}$K and $>10^{10}$g cm$^{-3}$, respectively. Subsequent decay of neutrons in beta decay $n\rightarrow p + e^{-} + \Tilde{\nu}_e$ is the ionization mechanism, in which the ionized particle starts its motion from the initial conditions of neutral geodesics. This scenario has recently been considered in the model of acceleration of ultrahigh-energy cosmic rays by a magnetized rotating Kerr black hole \cite{Tur-etal:2019:ApJ:}.

We assume that the disk is very slightly tilted with respect to the equatorial plane, which is orthogonal to the magnetic field lines at the poles. Studying the ionization mechanism and behavior of ionized particles near a black hole is important in the context of particle acceleration and energy extraction scenarios \cite{Par-Wag-Dad:APJ:1986:,Tur-Dad:2019:Universe:}, although here we restrict ourselves to the non-rotating neutral black hole case, where no energy is available for extraction. 
We consider a simple ionization model with conserved mechanical momenta of particles before and after ionization.  However, the canonical four-momenta of particles after ionization get additional contribution due to interaction with electromagnetic four-potential. Since in the combined magnetic field the only nonzero component of the four potential is $A_\phi$, the energy of a particle before and after ionization remains conserved, while angular momentum modifies. 
This model was proposed and studied in detail in \cite{Stu-Kol:2016:EPJC:,Kop-Kar:2018:APJ:} for the uniform magnetic field case as a particle acceleration scenario.

\begin{figure*}
\includegraphics[width=\hsize]{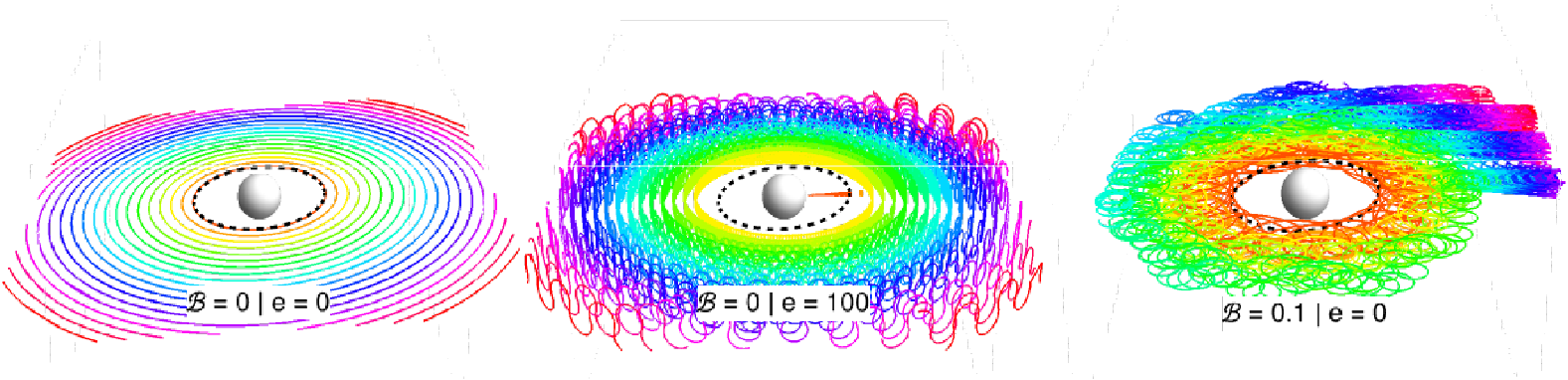}
\caption{Ionization of particles from thin Keplerian disk generated by a set of test particles on circular obit with different radii, here denoted by a different color. Before ionization, we can see just a slightly inclined ($\theta=1.5$) accretion disk (left figure). After the ionization in a pure BZ split-monopole (middle figure) the charged particles will follow a radial magnetic field creating a wave-like pattern, while after the ionization in a pure uniform MF (right figure), the particle will orbit a vertical magnetic field line creating curled trajectories around black hole, which is depicted as a gray sphere. The inner edge of the Keplerian disk (black dashed circle) is determined by the neutral test particle ISCO, $r=6$.}
\label{fig:IONintro}
\end{figure*}

A test particle inside the Keplerian accretion disk at the radius $r_0$ inclined with an angle $\theta_0$ with respect to the chosen equatorial plane has an energy and angular momentum  given by 
\beq
 \ce_{(\mJ)} = \frac{r_0-2}{\sqrt{r_0^2 -3r_0}}, \quad  \cl_{(\mJ)} = \frac{r_0 \sin\theta_0}{\sqrt{r_0-3}}. 
 \label{CandLinSCHW}
\eeq
Initial conditions for the particle are given by 
\bea
 x^\alpha &=& (t,r,\theta,\phi) =(0,r_0,\theta_0,0), \\
 u_\alpha &=& (u_t,u_r,u_\theta,u_\phi) =(\ce,0,0,\cl).
\eea
After ionization the particle starts to feel external magnetic field and angular momentum has to be modified as follows 
\beq
  \cl_{(\mD)} = \cl_{(\mJ)} + A_{\phi} (r_{0},\theta_{0}), \quad \ce_{(\mD)} = \ce_{(\mJ)}
\label{EL-ion}
\eeq
while energy remains unchanged since there is no scalar potential present in the chosen combined field solution in the Schwarzschild metric. 
Ionized particles from the Keplerian disk can now flow chaotically along magnetic field lines, oscillate around a circular orbit, or be caught by the black hole.

\begin{figure*}
\includegraphics[width=\hsize]{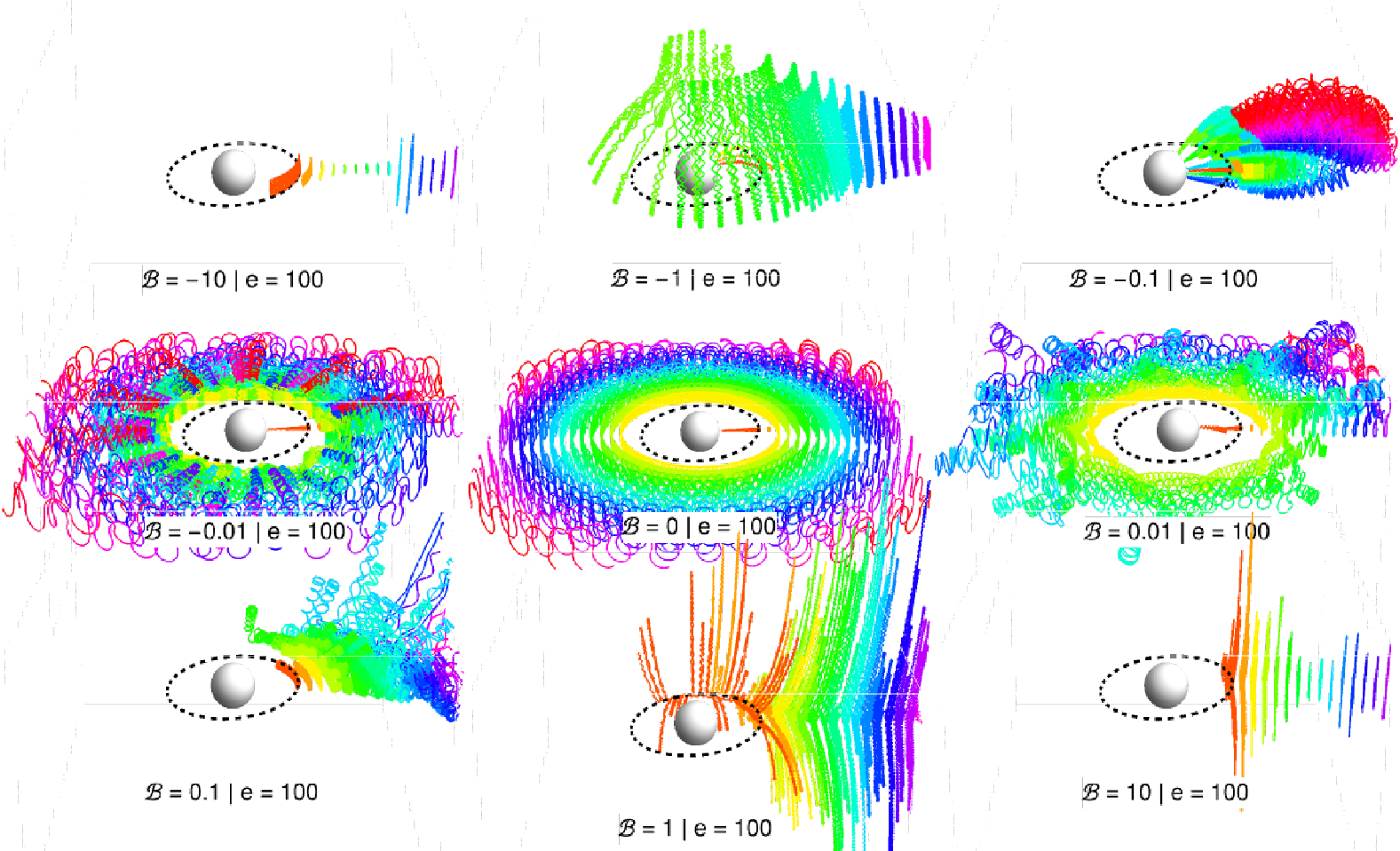}
\caption{
Evolution of thin Keplerian accretion disks around Schwarzschild BH immersed into the combined magnetic field with the BZ split-monopole parameter $e=\cb\beta=100$. Colors indicate particles with different initial conditions and are intended to better visualize individual trajectories. Initially, the particles follow circular geodesics with $\cl > 0$ and inclination from the equatorial plane $\theta_0=1.5$, while the magnetic field is directed along the vertical axis. The cases with $\cb < 0$ correspond to the looplike, while $\cb > 0$ to the jetlike magnetic field configurations. $\cb=0$ corresponds to the BZ split-monopole field without an external uniform magnetic field component. The black dashed circle indicates the ISCO for a neutral particle and the inner edge of the Keplerian accretion disk. One can notice that ionized particles in the jetlike magnetic field configurations tend to escape the black hole more than in the looplike configuration. 
\label{figION}
}
\end{figure*}

\begin{figure*}
\includegraphics[width=\hsize]{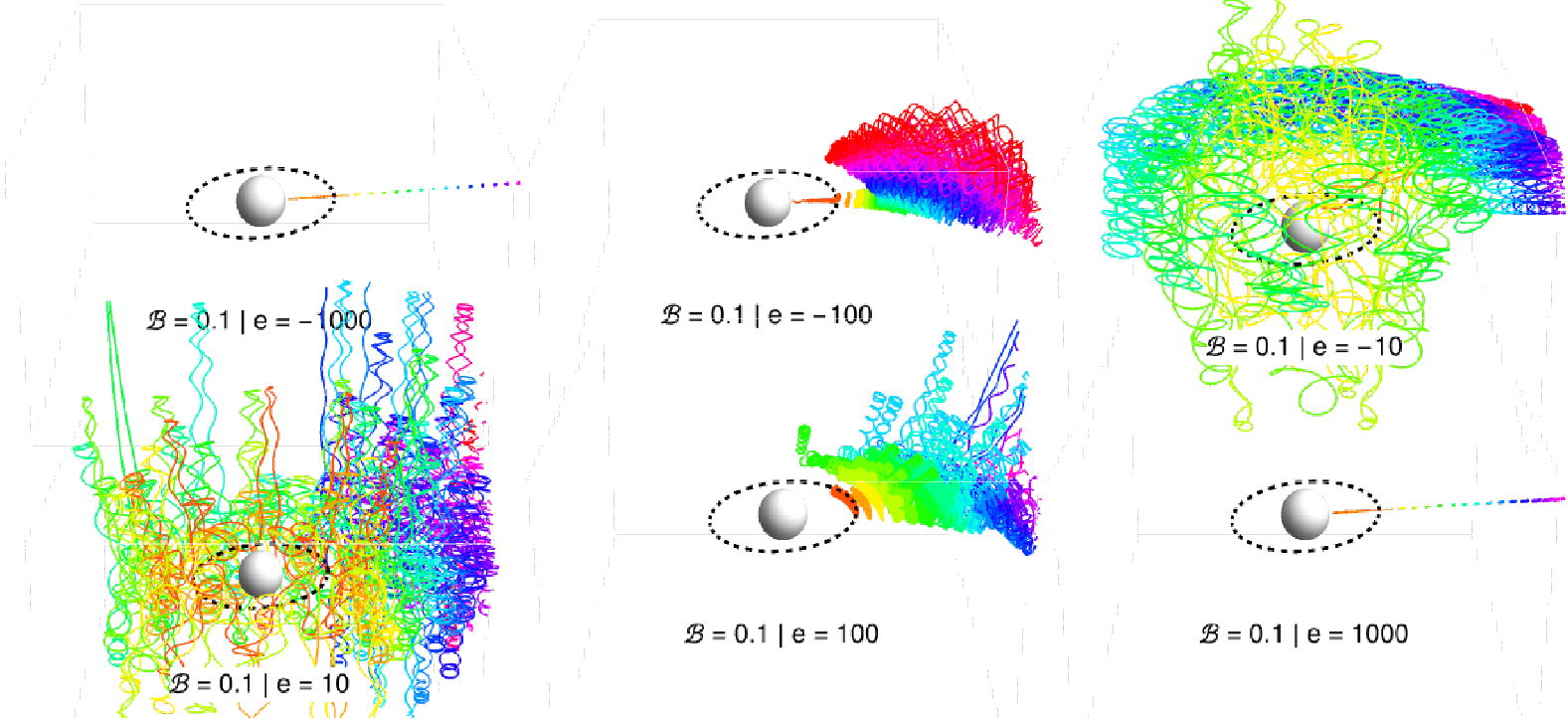}
\caption{ 
Ionization of a thin Keplerian disk for various BZ split-monopole parameter $e$. We keep the external uniform magnetic field the same $\cb=0.1$. For large values of the parameter $e=\pm1000$ we see that the ionized particle is just circling around the radial magnetic field line.  
See Figs. \ref{fig:IONintro} and \ref{figION} for additional information.
\label{figION2}
}
\end{figure*}


Considering the ionization of neutral particles at different points of the Keplerian accretion disk orbiting the Schwarzschild black hole in the presence of a combined magnetic field, we illustrate the behavior of ionized particles in Figs. \ref{fig:IONintro}, \ref{figION} and \ref{figION2}. In Fig.\ref{fig:IONintro} we demonstrate the particle ionization effect in both a pure BZ split-monopole and uniform magnetic field components, comparing them with the neutral particle motion. 
In Figs \ref{figION} and \ref{figION2}, since we agreed above to use $\cl > 0$, while the parameter $e$ is positive in Fig.\ref{figION} and negative in Fig.\ref{figION2}, due to symmetries discussed in Section \ref{sec:eff}, the two figures represent all four qualitatively different types of orbits differing in the direction of the Lorentz force, for each jetlike and looplike magnetic field configuration. 

Perhaps the most important observation from Fig.\ref{figION} is that regardless of the direction of the Lorentz force, the most plausible configuration for escaping charged particle trajectories is the jetlike magnetic field combination, specifically when $\beta \equiv e/B > 0$, i.e. when the two magnetic field components are aligned. Note that since the energy of an ionized particle is conserved according to (\ref{EL-ion}), the escape trajectories cannot extend to infinity. Despite that, simulations of particle ionization at different distances from black hole in a combined magnetic field show a tendency to escape trajectories only in the jetlike configuration. Such tendencies are likely caused by the combination of magnetic field lines since they were not observed in the past in similar settings for individual magnetic field configurations \cite{Stu-etal:2020:Universe:}. 

In contrast to the jetlike combination, in the looplike configuration, when the two magnetic field components are pointing in opposite directions, we see that the ionized particles tend to either fall into the black hole or oscillate in the black hole vicinity in a quasi-bounded state. In this combination, almost no escape trajectories are observed. Since the combined field lines form loops connecting the disk with the black hole, while charged particles follow magnetic field lines, one can conclude that the looplike configuration supports accretion of charged particles into the black hole.  

One can also note that the escaping tendency in the jetlike configuration is stronger when both field components are of comparable strengths. In other words, as shown in Fig.\ref{figION2} for very large $e$ and small $\BB$, the trajectories in the looplike configuration are similar to the looplike case, and no escaping tendency of charged particles is observed. This is simply due to the fact that in such cases the resulting field is mainly of a BZ split-monopole shape since the uniform component is too weak to considerably affect the combined field's shape. Therefore, a strong jetlike effect of charged particle trajectories requires the following condition 
\beq 
\cb > 1, \quad 1<\beta<10^3,  
\label{eq:jet-cond}
\eeq 
which is well full-filled in realistic astrophysical situations.

\begin{figure*}
\includegraphics[width=\hsize]{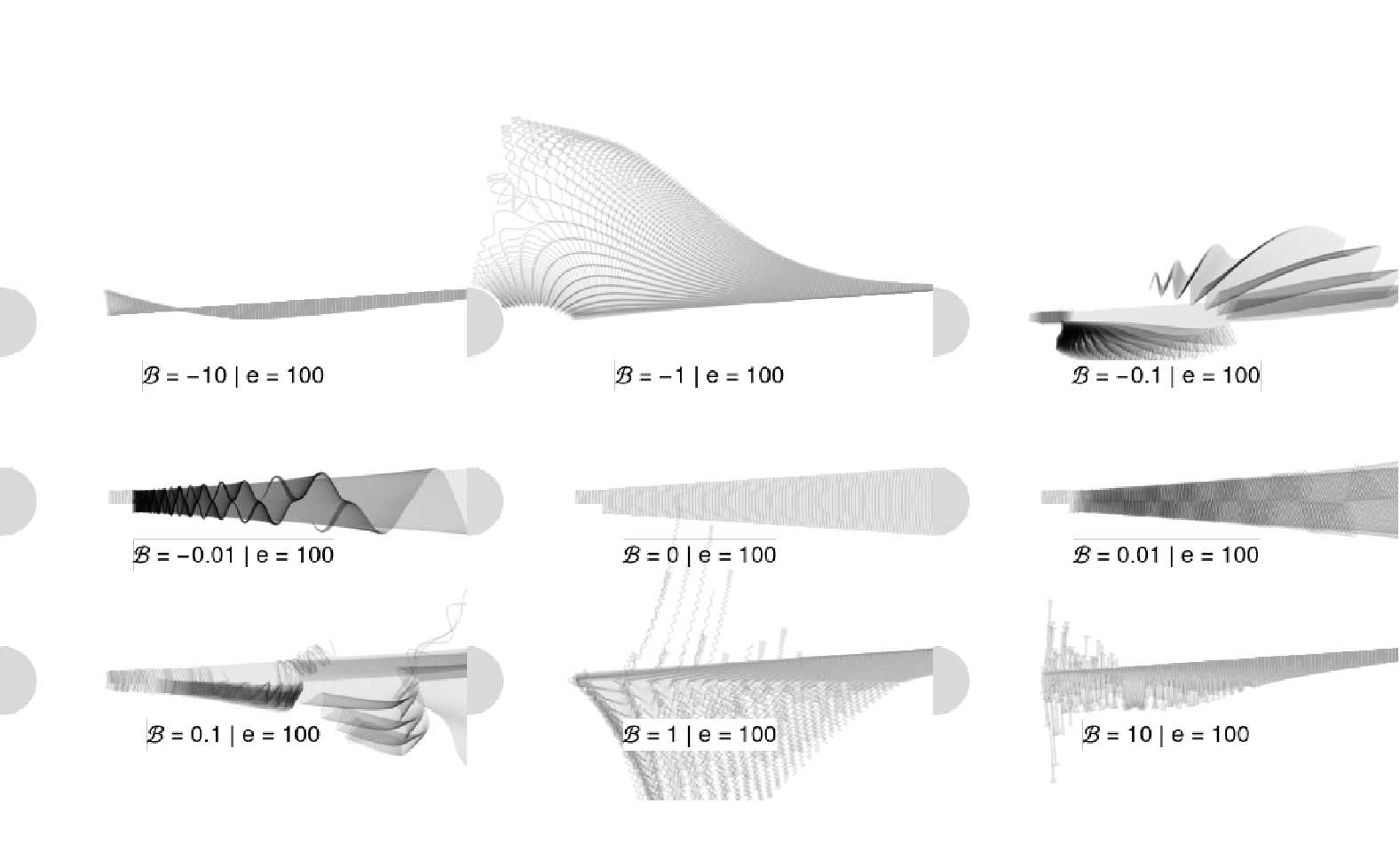}
\caption{Probability to find charged particles in combined black hole magnetosphere after the ionization of the Keplerian accretion disk. A visual representation of the density of particle trajectories in a restricted axially symmetric phase space is an analogue of Van Allen radiation belts for black holes. This figure uses the same initial conditions as in Fig.~\ref{figION}.
} \label{fig:belt1}
\end{figure*}

\begin{figure*}
\includegraphics[width=\hsize]{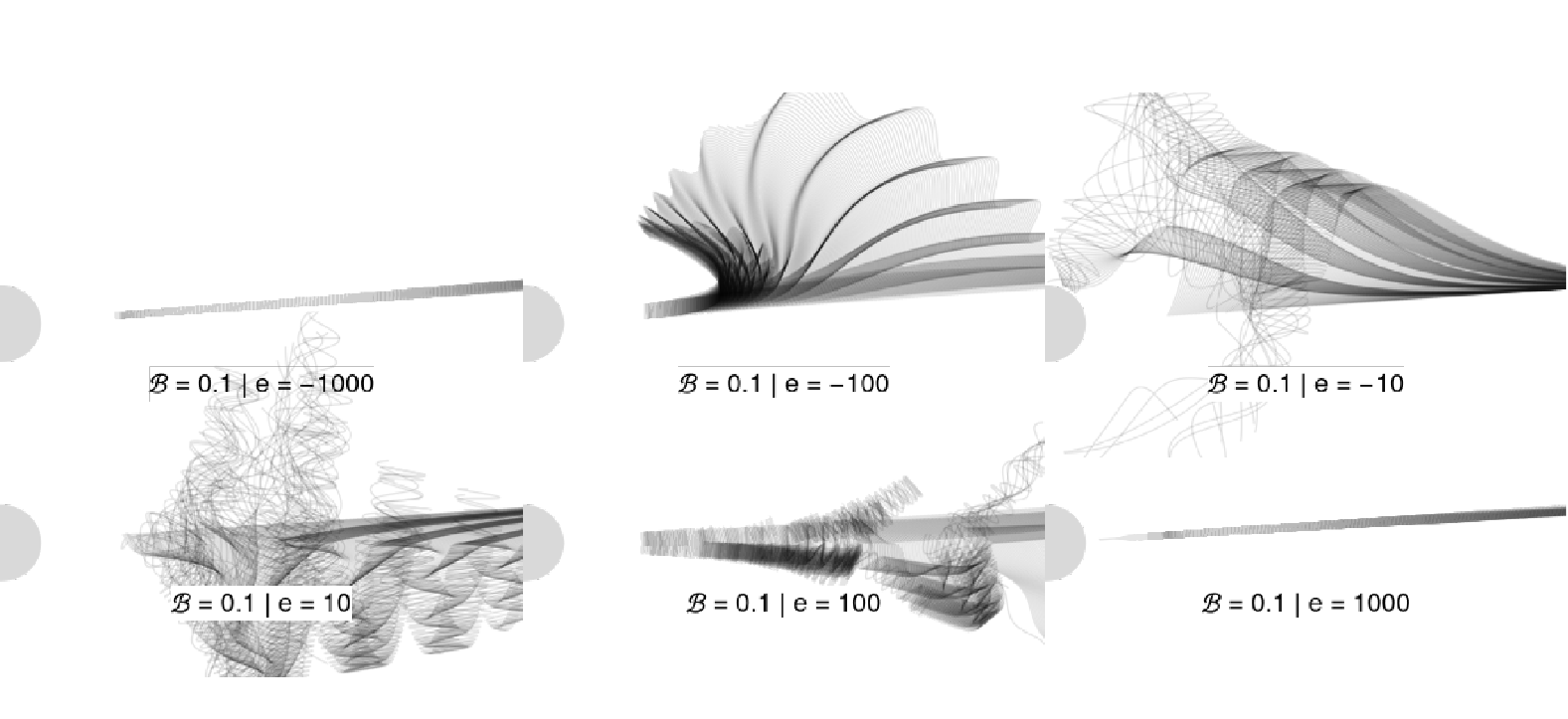}
\caption{
Probability to find charged particles in combined black hole magnetosphere after the ionization of the Keplerian accretion disk. A visual representation of the density of particle trajectories in a restricted axially symmetric phase space is an analogue of Van Allen radiation belts for black holes. This figure uses the same initial conditions as in Fig.~\ref{figION2}.
} \label{fig:belt2}
\end{figure*}

The process of neutral particle ionization can be considered as a special case of a more general process of particle scattering, where the particle's momenta are changed along with its charge.  
In this section, we aimed to explore whether the ionized particles have preferred types of motion in the combined magnetic field and whether the particle trajectories form any structures in the black hole's neighborhood, where the ionized particles are more likely to be found. Such structures, full of ionized particles, are plotted in Figs.~\ref{fig:belt1} and \ref{fig:belt2} and in some sense remind of the Van Allen radiation belt formed in the Earth's dipole magnetic field.

As mentioned above, the use of the Schwarzschild metric does not allow the charged particles to gain energy through the considered ionization mechanism. Therefore, it is interesting to study similar processes in a combined magnetic field around a rotating Kerr black hole, where the energy extraction mechanisms can take place. However, we leave these studies in the Kerr case to future work.

\section{Summary \& conclusion}

The structures of magnetic fields around astrophysical black holes have not yet been properly resolved. In some regions with relatively higher mass density, the magnetic field is expected to have a quite complex character due to turbulent processes inside the surrounding plasma. Yet, in predominantly dilute regions, like those outside an accretion disk, the field lines can be of regular and even extended large-scale shape, which is supported by polarimetric observational studies related to various astrophysical black hole candidates and their relativistic jets \cite{Nak-etal:2018:APJ:}. However, even in those cases, modeling realistic magnetic field lines in a strong gravity regime remains challenging. Usually, the black hole magnetosphere is modeled by a single magnetic field source, either related to an accreting plasma or to the field of external origin (e.g. that of a companion neutron star, Galactic magnetic field, etc.). Nevertheless, it also seems natural that the final form of the magnetosphere is governed by a combination of both internal and external fields of different origins.

Here, we propose for the first time a simple analytic model for a black hole magnetosphere formed by a combination of two independent magnetic fields. As an internal and therefore stronger magnetic field component we choose the Blandford-Znajek split-monopole magnetic field, which is a solution to force-free electrodynamics but it is also supported by GRMHD and GRPIC simulations \cite{Tch-Nar-McK:2010:APJ:,Par-Phi-Cer:2019:PRL:}. As an external, weaker magnetic field component, we choose the asymptotically homogeneous magnetic field given by the Wald solution, which can be effectively used to model large-scale magnetic fields within a limited region of space. Close to the black hole, the magnetosphere is dominated by the BZ split monopole component, slightly transforming into a uniform field with increasing distance from the black hole. In order to see the primary effects of combined fields in a regime of strong gravity, we used the Schwarzschild metric. 

In order to simplify the problem, we restricted ourselves to an axially symmetric situation when the two field components are either aligned in the same or in opposite directions. We have shown that when the magnetic field components are aligned, the resulting field is of a paraboloidal shape, reminding the collimated jet structures at the launching regions. We call this combination jetlike. Indeed, as we have shown in Section \ref{sec:Disk}, this is an optimal configuration for escaping charged particle trajectories among the considered individual fields and their combinations. The jetlike combination of magnetic field components creates zero-potential lines of paraboloidal structure, which is equivalent to the regions of the minimal potential energy. Therefore, close to these lines, extending from the black hole to infinity, charged particles may have better chances to escape.  

In the case where the magnetic field components are directed in opposite directions, the combined field lines create magnetic loops connecting the equatorial accretion disk region with the black hole and the magnetic null points at the polar axis.  
We call this combination looplike.  
Magnetic null points that exist only in the looplike combination can be relevant when plasma is injected into these points, as it could lead to magnetic reconnection and acceleration of particles \cite{Kar-Kop:2009:CQG:,2013IJAA....3...18K}. 
We have shown that charged particles in the looplike configuration are more likely bounded into the region in the black hole vicinity, and almost no particles escape to infinity in this configuration in the form of jets, unless the special choice of initial conditions. This effect of the absence of the jets due to interplay between the external and internal magnetic field components of opposite directions can potentially serve as a possible explanation of the radio-quiet AGNs, which are known to have no large-scale relativistic jets \cite{2019NatAs...3..387P}. This in fact can distinguish them from the radio-loud AGNs with relativistic jets, if the external magnetic field component is oriented in the same direction as the internal one. The detailed analysis of the role of external magnetic field components on the observables of radio-quite and radio-loud AGNs is left to future work. 

The BZ split-monopole magnetic field solution needs supporting electric currents floating in the equatorial plane and such currents generate dipole magnetic moment which could be assigned to a split magnetic monopole located inside the black hole. The split-monopole magnetic moment will be oriented in alignment with the external uniform magnetic field (jetlike configuration) or against it (looplike configuration). One can assume that the magnetically aligned jetlike configuration is energetically more favorable.

Based on the study of the effective potential, we performed a detailed analysis of different types of charged particle motion in the two new configurations, mapping stable circular orbits for possible particle accumulation. The minima of effective potential can be filled up by ionized particles trapped in the black hole vicinity. We investigated the trajectories of charged particles numerically and classified them.  We have shown that the dynamics of charged particles can be complicated and chaotic if the Lorenz force acting on the particles is of a magnitude comparable to the gravitational attraction of the black hole. In a stronger magnetic field, the particles closely follow the magnetic field lines. We also found that stable circular orbits can exist outside the equatorial plane in both jetlike and looplike combinations. These orbits cannot exist in a scenario involving only a pure uniform magnetic field. At the equatorial plane, the dynamics of charged particles are governed by the uniform magnetic field component, as the BZ split-monopole component vanishes at the equatorial plane.

We demonstrated that ionized particles from the Keplerian accretion disk can escape to infinity along the magnetic field lines, providing the energetic condition for escaping trajectories. This creates a stable and well-collimated flow of charged particles along the vertical $z$-axis. If the energy of charged particles is less than the critical escape energy, the particles remain in trapped states around the black hole, forming radiative belts of various shapes depending on the ratio between the BZ split-monopole and uniform MF components. The portion of the ionized particles will also fall into the BH. 

From multiple numerical experiments presented in this paper, we conclude that the jetlike combination of the field components encourages the charged particles to escape to infinity. This effect is not as strong in the cases of considered individual magnetic field components. Therefore, combining the BZ split-monopole and weak uniform magnetic field components can be a better and more plausible setup for the jet acceleration scenarios. On the other hand, if the field components are aligned in opposite directions, then such a configuration supports accretion of ionized particles from the accretion disk into the black hole, as more particles are captured by the black hole. Although the BZ split-monopole component in the black hole vicinity is expected to be stronger than the external uniform magnetic field, the optimal jetlike acceleration is achieved when the condition (\ref{eq:jet-cond}) holds. If we assume an external field to be of the order of a few G, then for the motion of protons and ions the internal field component's strength has to be $<10^3$G. For electrons, the internal field can be stronger by the mass ratio between protons and electrons. As one can see, condition (\ref{eq:jet-cond}) is quite realistic and applicable to both stellar-mass and supermassive black holes, which makes the proposed jetlike magnetosphere astrophysically plausible.

\begin{acknowledgements}
This work was partially supported by the Ministry of Education and Science (MES) of the Republic of Kazakhstan (RK), Grant No. AP13067667, and the Czech Science Foundation Grant (GA\v{C}R) No.~\mbox{23-07043S}. 
S.T. acknowledges the support through the postdoctoral fellowship program of Al-Farabi Kazakh National University.  
A.T. acknowledges the Alexander von Humboldt Foundation for its Fellowship. 
\end{acknowledgements}



\providecommand{\noopsort}[1]{}\providecommand{\singleletter}[1]{#1}%

\end{document}